\documentclass[final]{siamltex}

\usepackage{amsmath}
\usepackage{amsfonts}
\usepackage{amssymb}
\usepackage{graphics}
\usepackage{graphicx}
\usepackage{amsbsy}
\usepackage{exscale}
\newtheorem{remark}{{\it Remark\/}}[section]

\newcommand{\E}{{\mathbb E}}

\newcommand{\T}{{\mathbb T}}
\newcommand{\R}{{\mathbb R  }}
\newcommand{\eps}{{\epsilon  }}
\newcommand{\summ}[1]{\sum_{#1 =1}^{\infty}}
\newcommand{\normm}[1]{\mathbb{E}\left( \sup_{0 \leq t \leq T} \left\| #1
                       \right\|^2 \right)}
\newcommand{\nrm}[2]{\mathbb{E}\left( \sup_{0 \leq t \leq T} \left\| #1
                       \right\|^{2 #2} \right)}

\title{ANALYSIS OF WHITE NOISE LIMITS FOR STOCHASTIC SYSTEMS WITH TWO
      FAST RELAXATION TIMES\thanks{This work was supported by the Engineering and Physical 
      Sciences Research Council.}}

\author{G. A. Pavliotis\footnotemark[3] 
\thanks{Current Address: Department of Mathematics, Imperial College 
London, London SW7~2AZ, UK (g.pavliotis@imperial.ac.uk) }
\and A. M. Stuart\thanks{Mathematics Institute, Warwick University,
Coventry, CV4~7AL, UK (pavl@maths.warwick.ac.uk, stuart@maths.warwick.ac.uk).}
}
       
\begin{document}

\maketitle

\begin{abstract}
In this paper we present a rigorous asymptotic analysis for stochastic systems with
two fast relaxation times. The mathematical model analyzed in this paper consists of
a Langevin equation for the particle motion with time-dependent force
constructed through an infinite
dimensional Gaussian noise process. We study the limit as the particle relaxation
time as well as the correlation time of the noise tend to zero and we obtain the
limiting equations under appropriate assumptions on the Gaussian noise. We show that
the limiting equation depends on the relative magnitude of the two fast time scales
of the system. In particular, we prove that in the case where the two relaxation
times converge to zero at the same rate there is a drift correction, in addition to
the limiting It\^{o} integral, which is not of Stratonovich type. If, on the other
hand, the colored noise is smooth on the scale of particle relaxation then the drift
correction is the standard Stratonovich correction. If the noise is rough on this
scale then there is no drift correction. Strong (i.e. pathwise) techniques are used
for the proof of the convergence theorems.
\end{abstract}



\pagestyle{myheadings}
\thispagestyle{plain}
\markboth{G.A. PAVLIOTIS AND A.M. STUART}{WHITE NOISE LIMITS FOR SYSTEMS WITH TWO
      FAST RELAXATION TIMES}

\section{Introduction}
\label{sec:intro}

Many physical systems are subject to either additive or multiplicative noise. The
dynamics of such systems are quite often adequately described by systems of
stochastic differential equations. There are various applications where the noise in
the physical system under investigation has a non--trivial spatio--temporal
structure and where it is not realistic to model it is as a white noise process. The term
colored noise is used for such a noise process.

It is a well known result that if we approximate white noise by a smooth, colored
process, then at the limit as the correlation time of the approximation tends to
zero, the smoothed stochastic integral converges to the Stratonovich stochastic
integral
\cite{blakenship, wong}, \cite[ch.
10]{arnold_l}. To be precise, consider the SDE (written here in one dimension for simplicity)
\begin{equation}
\dot{x} = b(x) + \frac{f(x) \eta(t/\epsilon^2)}{\epsilon},
\label{e:col_intro}
\end{equation}
where $b(x), \, f(x)$ are Lipschitz continuous and $\eta(t)$ is a continuous mean
zero Gaussian process with $\E (\eta(t) \eta(s)) = \frac{1}{2}e^{- |t - s|}$ . Then,
the results of \cite{blakenship, wong} imply that, as $\epsilon$ converges to $0$,
the solution of \eqref{e:col_intro} converges weakly to $X(t)$ which satisfies\footnote{Throughout
the paper we will use the notation $\int_0^t f(x(s)) \, d \beta(s)$ (respectively $ f(x(t))
\dot{\beta}(t)$) to denote the It\^{o} stochastic integral (respectively differential) and
$\int_0^t f(x(s)) \circ \, d \beta(s)$ (respectively $ f(x(t)) \circ \dot{\beta}(t)$) for the
Stratonovich stochastic integral (respectively differential). Furthermore, we will refer to an
It\^{o} or Stratonovich SDE depending on how we choose to interpret the stochastic integral in the
equation.}
\begin{equation}
\dot{X}(t) = b(X) + \frac{1}{2}f(X) f'(X) + f(X) \dot{\beta}.
\label{e:strat_intro}
\end{equation}
Here $\beta(t)$ denotes a standard one--dimensional Brownian motion. The term
$\frac{1}{2} f f'$ is sometimes referred to as the Stratonovich correction. This
result has been extended in various ways, including the case of multiple It\^{o}
integrals \cite{kushner1} as well as linear \cite{bouc, kushner2, carmona_2,
fannjiang} and semilinear SPDE, \cite{zabczyk} and the references therein. Moreover,
the case of infinite dimensional noise has also been studied \cite{carmona_2,
dowell,fannjiang}. In the context of the theory of turbulent diffusion the results of the
aforementioned papers are concerned with convergence of rapidly decorrelating in
time velocity fields to the Kraichnan model for passive tracers \cite{kraichnan}.

The main interest of this work is to extend these results to situations where
inertial effects are taken into account. For motivation, consider the motion of a
particle with relaxation time $\tau$ under the influence of a force field $b(x)$ and
subject to dissipation and colored multiplicative noise:
\begin{equation}
\tau \ddot{x} = b(x) - \dot{x} + \frac{f(x) \eta(t/\epsilon^2)}{\epsilon}.
\label{e:lang_in}
\end{equation}
We are interested in analyzing the limit of \eqref{e:lang_in} as both $\tau$ and
$\epsilon$ tend to $0$. It should be expected that these two limits do not commute.
To see this, first let $\epsilon \rightarrow 0$ while keeping $\tau$ fixed to obtain
the SDE \footnote{It is easy to check that in this case there is no Stratonovich
correction to the It\^{o} integral, because of the regularity of $x$ .}
$$
\tau \ddot{x} = b(x) - \dot{x} + f(x) \dot{\beta}.
$$
Taking now the limit as $\tau \rightarrow 0$ leads to the It\^{o} SDE \cite[ch. 10]{nelson}
$$
\dot{x} = b(x) + f(x) \dot{\beta}.
$$
On the other hand, if we first take the limit as $\tau \rightarrow 0$ and then let $\epsilon
\rightarrow 0$, we end up with equation
\eqref{e:strat_intro}.

Because of this lack of commutativity, it not clear what the limiting equation
should be as we let $\epsilon$ and $\tau$ both tend to $0$ at the same time. This is
the sometimes referred to as the {\it It\^{o}--versus--Stratonvich problem} in the
physics literature \cite{reimann}. The correct form of the limiting SDE and, in
particular, the presence or otherwise of a drift correction term in addition to the
limiting It\^{o} integral--the {\it noise induced drift}-- is of particular
importance in the theory of Brownian motors \cite{reimann}, noise induced phase
transitions \cite{sancho, lefever, mangioni} and the dynamics of fronts
\cite{sancho1}.

The purpose of this paper is to investigate the It\^{o} versus Stratonovich problem
rigorously. Let us discuss now the main results of this paper in the one dimensional
setting. The $1d$ version of the model considered in this paper reads
\begin{subequations}
\begin{equation}
\tau_0 \epsilon^{\gamma} \ddot{x} = b(x) - \dot{x} + \frac{f(x)
\eta(t/\epsilon^2)}{\epsilon},
\label{e:lang_ou}
\end{equation}
\begin{equation}
\dot{\eta} = - \alpha \eta  + \sqrt{\lambda} \dot{\beta},
\label{e:ou_intro}
\end{equation}
\label{e:syst_intro}
\end{subequations}
where $\alpha, \, \lambda, \, \tau_0$ are positive $\mathcal{O}(1)$ parameters and
$\gamma \in (0,\infty)$. In this paper we show that three possible limiting
equations result, depending on the magnitude of the particle relaxation time
relative to that of the noise correlation time, i.e. on the exponent $\gamma$. In
particular, for $\gamma \in (0,2)$ we show that the limiting equation is the It\^{o}
SDE
\begin{equation}
\dot{X}(t) = b(X)  + \frac{\sqrt{\lambda}}{\alpha} f(X)
\dot{\beta}.
\label{e:lim1_intro}
\end{equation}
For $\gamma \in (2, \infty)$ we obtain, at the limit $\epsilon \rightarrow 0$, the
Stratonovich SDE
\begin{eqnarray}
\dot{X}(t) & = &  b(X) +  \frac{\sqrt{\lambda}}{\alpha} f(X) \circ \dot{\beta}
         \nonumber \\ & = &
 b(X) + \frac{\lambda}{2 \alpha^2}f(X) f'(X) + \frac{\sqrt{\lambda}}{\alpha} f(X)
\dot{\beta},
\label{e:lim2_intro}
\end{eqnarray}
For $\gamma = 2$ the limiting SDE can be interpreted in neither the It\^{o} nor the
Stratonovich sense; we obtain:
\begin{equation}
\dot{X}(t) = b(X) + \frac{\lambda}{2 \alpha^2 (1 + \tau_0 \alpha)}f(X) f'(X) +
\frac{\sqrt{\lambda}}{\alpha} f(X) \dot{\beta},
\label{e:lim3_intro}
\end{equation}
Let us make some remarks concerning \eqref{e:lim3_intro}. If we define the
stochastic integral
\begin{equation}
\int_0^t f(X(s)) \widehat{\circ} \, d \beta(s) := \frac{\lambda}{2 \alpha^2 (1 +
\tau_0 \alpha)} \int_0^tf(X(s)) f'(X(s)) \, ds + \frac{\sqrt{\lambda}}{\alpha}
\int_0^t f(X(s)) \, d \beta(s),
\label{e:int_intro}
\end{equation}
then this integral obeys neither the It\^{o} nor the Statonovich (i.e.
Newton--Leibnitz) calculus. Let us now define the stochastic integral in
\eqref{e:int_intro} as the limit of Riemann sums
\begin{equation}
\int_0^t f(X(s)) \widehat{\circ} \, d \beta(s) \approx  \frac{\sqrt{\lambda}}{\alpha}
\sum_{j =1}^N \left( \mu f(X(s_j)) + (1 - \mu) f(X(s_{j -  1}) \right) ( \beta(s_j) -
\beta(s_{j - 1}) ),
\nonumber
\end{equation}
with $\mu \in [0,1]$. The stochastic integral \eqref{e:int_intro} corresponds to the
choice $\mu = \frac{1}{2 (1 + \tau_0 \alpha)}$. As is well known, e.g.
\cite{oksendal}, the It\^{o} integral corresponds to the choice $\mu = 0$, whereas
the Stratonovich integral to $\mu = \frac{1}{2}$. Notice that letting $\tau_0$ in
\eqref{e:lim3_intro} vary in $(0,\infty)$ interpolates between these two well known
integrals.

The one dimensional model \eqref{e:syst_intro} was studied by Graham and Schenzle in
\cite{graham} using formal singular perturbation analysis for the corresponding
Fokker--Planck equation in the spirit of \cite{blakenship}. Similar questions to the
one studied in this paper were investigated by Givon and Kupferman in \cite{raz} for
SDE limits of discrete dynamical
 systems
with scale separation. Specific examples were presented where the limiting SDE are
neither of It\^{o} nor of Stratonovich type. A formal derivation of the results
reported in this paper together with extensive numerical simulations were presented
in \cite{paper3_stuart}.

In this paper we base our rigorous derivation of the limiting SDE for the infinite
dimensional version of \eqref{e:syst_intro}--see equation below-- using the pathwise
techniques developed in \cite{paper1_stuart}, following the work of \cite{dowell}.
Our method enables us to treat the infinite dimensionality of the noise in a rather
straightforward way and, in addition, to prove strong convergence results.
Furthermore, we are able to prove upper bounds on the convergence rate in
$L^p(\Omega ; C([0,T]; \R))$. The numerical results reported in \cite{paper3_stuart}
indicate that the upper bounds are in fact sharp.

This paper is organized as follows. In section \ref{s:res} we describe the equations
that we will be studying and we present the convergence theorems. In section
\ref{sec:prelim_lemmas} we present various preliminary results which are necessary
for the proof of our convergence theorems. In section \ref{sec:ito_vs_strat} we show
that the structure of the limiting equations depends crucially on $\gamma$. Our
convergence theorems are proved in section \ref{sec:converg_thm}. In section
\ref{sec:applications} we present two applications of the convergence theorems, with
particular emphasis on the inertial particles problem considered in \cite{
paper1_stuart, inertial_1, inertial_2}. Finally, section \ref{sec:conclusions} is
devoted to some concluding remarks.
%
%
%
%
\section{Description of the Model and Statement of Main Results}
\label{s:res}

In some of the applications of interest to us the driving colored noise is infinite
dimensional. This arises, for instance, in Gaussian random field models of
turbulence such as those pioneered by Kraichnan \cite{kraichnan}, and
generalizations to include noise correlation times \cite{carmona_2, carmona,
fannjiang,kramer_JSP,inertial_1, inertial_2}. Such applications are described in
section \ref{sec:applications}. In this section we formulate the problem for
infinite dimensional driving noises, and state our main results.
\subsection{The Model}
\label{subsec:intro_model}
We consider the Langevin dynamics for a particle moving in $\mathbb{R}^d, \, d \geq
1$ under the influence of a forcing term $b(x)$ and a rapidly decorrelating in time
random field $v(x,t)$:
\begin{equation}
\epsilon^{\gamma}\tau_0 \ddot{x} = b(x) + \frac{v(x, t/\eps^2)}{\epsilon} - \dot{x},
\; \; x \in \mathbb{R}^d,
\label{eqn_motion_case_3}
\end{equation}
where $\gamma \in (0, \infty)$ and $\epsilon \ll 1$. The field $v(x,t)$ is a
generalized Ornstein--Uhlenbeck process. This is a mean zero, Gaussian process which
can be constructed as the solution of the vector valued SPDE
\begin{equation}
d v = - \widehat{A} v \, dt + d \widehat{W}.
\label{e:ou_gen}
\end{equation}
Here we take $\widehat{A}: D(\widehat{A}) \rightarrow (L^2(\Omega))^d $, where
$\Omega \subset \R^d$ and $\widehat{W}$ is a $\widehat{Q}$--Wiener process on $H =
(L^2(\Omega))^d$. We assume that $\widehat{A}$ is a strictly positive self--adjoint
operator on the Hilbert space $H$ and that, furthermore, it has the same
eigenfunctions $\{f_k \}_{k =1}^{\infty}$ as $\widehat{Q}$:
$$
\widehat{A} f_k = \alpha_k f_k, \qquad \widehat{Q} f_k = \lambda_k f_k.
$$
We now assume that there exist vectors $h_k \in \R^d$ and positive definite selfadjoint
operators $A, \, Q$ on $L^2(\Omega)$ such that
$$
f_k = h_k \phi_k, \qquad A \phi_k = \alpha_k \phi_k, \qquad Q \phi_k = \lambda_k
\phi_k.
$$
Using this we can write

\begin{equation}
v(x,t) = f(x) \eta(t) = \summ{k} h_k \phi_k(x) \eta_k(t),
\nonumber
\end{equation}
where $\eta(t) :\ell_2 \rightarrow \R$ is defined through the equation
\begin{equation}
d \eta = - A \eta  \, dt + d W.
\label{eqn:ou_k}
\end{equation}
Here, abusing notation, we have used $A, \, Q \in L( \ell_2)$ with
$$
A = \mbox{diag} \{ \alpha_k \}, \qquad Q = \mbox{diag} \{ \lambda_k \}.
$$
Furthermore, $W$ is a $Q$--Wiener process on $\ell^2$:
$$
W(t) = \summ{k} \sqrt{\lambda_k} e_k \beta_k(t),
$$
with $\{ e_k \}_{k =1}^{\infty} $ being the standard basis in $\ell_2$ and
$\beta_k(t)$ mutually independent standard one--dimensional Brownian motions. We
remark that, for each fixed $x$, $f$ is a linear operator from $\ell_2$ to $\R^d$:
$f \in L(\ell^2, \R^d)$.

Using now the fact that $\beta(ct) = \sqrt{c} \beta(t)$ in law, we can finally write
our model in the following form:
\begin{subequations}
\begin{equation}
\eps^{\gamma} \ddot{x} = b(x) + \frac{v(x,t)}{\eps} - \dot{x},
\label{e:motion_1}
\end{equation}
\begin{equation}
v(x,t) = f(x) \eta(t),
\label{e:vel_1}
\end{equation}
\begin{equation}
d \eta = - \frac{1}{\eps^2}A \eta  \, dt + \frac{1}{\eps} d W.
\label{e:ou_2}
\end{equation}
\label{e:main}
\end{subequations}
To simplify the notation we have set $\tau_0 =1$ in \eqref{e:motion_1}. In the
sequel we will use the both notations $v(x,t)$ and $f(x) \eta(t)$ for the random
field.
\subsection{Statement of Main Results}
\label{subsec_intro_results}
Our goal now is to obtain the limiting equations of motion, as $\epsilon \rightarrow
0$. In order to prove our convergence theorems we will need to impose various
conditions on the spectrum of the Wiener process, the eigenvalues of the operator
$A$, the eigenfunctions $\{\phi_k(x) \}_{k =1}^{\infty}$ and the drift term
$b(x)$. The conditions that we have to impose are more severe for $\gamma \geq 2$,
since in this parameter regime we will need more integrations by parts in order to
obtain the limiting equations.

We will use the notation $\| \cdot \|$ to denote the Euclidean norm in
$\mathbb{R}^d$. Subscripts with commas will be used to denote partial
differentiation.

As has already been mentioned, we take $A$ to be a self--adjoint, positive operator on
$L^2(\Omega)$. We assume that the eigenvalues $\{ \alpha_k \}_{k =1}^{\infty}$ of $A$ satisfy
\begin{equation}
\dots \geq \alpha_{k+1} \geq \alpha_k \geq \omega > 0,  \qquad  \summ{k}
\frac{\lambda_k}{2 \alpha_k} < \infty.
\label{e:eigs_pos}
\end{equation}
The eigenfunctions of $A$ are normalized so that their $L^2(\Omega)$ norm is set to
$1$: $\|\phi_k \|_{L^2(\Omega)} =1$. Moreover, for $\gamma \in (0,2)$ we assume that
there exist constants $C >0, \, \alpha, \, \beta $ such that
\begin{eqnarray}
    \left\{ \begin{array}
            {r@{\quad} l}
            \phi_k(x) \in C^2_b(\Omega), \; \; k =1,2, \dots,
             \medskip

               \\
             \|\phi_k(x) \|_{L^{\infty}(\Omega)} \leq C \, \alpha_k^{\alpha},
         \, \|D \phi_k(x) \|_{L^{\infty}(\Omega)} \leq C \, \alpha_k^{\beta}.
%
%
%
            \end{array}   \right.
\label{eqn:conds_efs_g_s}
\end{eqnarray}
The conditions for $\gamma \geq 2$ are more severe. We assume that there exist
constants $C >0, \, \alpha, \, \beta, \,\gamma, \, \delta$ such that
\begin{eqnarray}
    \left\{ \begin{array}
            {r@{\quad} l}
            \phi_k(x) \in C^3_b(\Omega), \; \; k =1,2, \dots,
             \medskip
               \\
             \|\phi_k(x) \|_{L^{\infty}(\Omega)} \leq C \, \alpha_k^{\alpha},
         \, \|D \phi_k(x) \|_{L^{\infty}(\Omega)} \leq C \, \alpha_k^{\beta},
             \medskip
               \\
             \|D^2 \phi_k(x) \|_{L^{\infty}(\Omega)} \leq C \,
         \alpha_k^{\gamma},
         \, \|D^3 \phi_k(x) \|_{L^{\infty}(\Omega)} \leq C \,
         \alpha_k^{\delta}.
            \end{array}   \right.
\label{eqn:conds_efs_g_b}
\end{eqnarray}
\begin{remark}
\label{rem:ell1}
At this level of generality and, in particular, since we do not make any specific
assumptions on the operator $A$, we do not have any detailed information on the
$L^{\infty}$ norm of the eigenfunctions $\{ \phi_k\}_{k =1}^{\infty}$ and their
derivatives. Much is known when $A$ is a uniformly elliptic operator, see e.g.
\cite[Ch. 5]{sogge}, \cite{aurich, grieser} and the references therein. In
particular, the results from \cite{grieser} imply that, when $A$ is a uniformly
elliptic operator with smooth coefficients and Dirichlet or Neumann boundary
conditions on some bounded domain $\Omega \subset \R^d$ with smooth boundary, then
the following estimate holds:
\begin{equation}
\|D^n \phi_k \|_{L^{\infty}(\Omega)} \leq C \alpha_k^{\frac{d-1 + \frac{n}{2}}{4}},
n = 0, 1, \dots
\label{e:einf_ell}
\end{equation}
\end{remark}

We will assume that the drift $b(x)$ is Lipschitz continuous:
\begin{equation}
\| b(x) - b(y) \| \leq C \|x - y \|, \quad x,y \in \R^d.
\label{e:b_cond}
\end{equation}
Moreover, we will assume that there exist constants $C, \, r$ such that
\begin{equation}
\| h_k \| \leq C \, |\alpha_k|^{r}, \; \; k =1,2, \dots, .
\label{cond_coeff_h}
\end{equation}
Now we are ready to present the conditions that we have to impose on the spectrum of
the Wiener process. First, we need to ensure the existence and uniqueness of the
equations of motion \eqref{e:motion_1}. To this end, we assume that the velocity
field is sufficiently regular:\footnote{A simple variant of \cite[Thm.
5.20]{prato92} yields that $v(x,t) \in (C(\R^{+},C^1(\Omega)))^d$ provided that
there exists a $\zeta \in (0,1)$ such that
\begin{equation}
\summ{k} \lambda_k\alpha_k^{2(r + \alpha)- 1 - \zeta} < \infty,
\nonumber
\end{equation}
\begin{equation}
\summ{k} \lambda_k \alpha_k^{2(r + \zeta \gamma) -1 } < \infty.
\nonumber
\end{equation}
However, these conditions are not optimal and so we simply assume \eqref{eqn:conds_lip}.}
\begin{equation}
v(x,t) \in (C(\R^{+},C^1(\Omega)))^d.
\label{eqn:conds_lip}
\end{equation}
Assumption \eqref{eqn:conds_lip}, together with assumption \eqref{e:b_cond} ensure
that there exist almost surely a unique solution of the equations of motion
\eqref{e:motion_1}, when the initial conditions for \eqref{e:ou_2} are distributed
according to the invariant measure of this process.
Furthermore, for $\gamma < 2$ we have to assume conditions of the form:
\begin{subequations}
\begin{equation}
\summ{k} \sqrt{\lambda_k}\alpha_k^{(r + \alpha - \frac{1}{2} -\rho)}
< \infty,
\label{eqn:conds_g_s_1}
\end{equation}
\begin{equation}
\summ{k} \sqrt{\lambda_k}\alpha_k^{(r + \beta - \frac{1}{2} -\rho)}
< \infty,
\label{eqn:conds_g_s_2}
\end{equation}
\label{eqn:conds_g_s}
\end{subequations}
The specific value of the exponent $\rho$ will be given when stating our convergence
theorems. For $\gamma \geq 2$, in addition to the \eqref{eqn:conds_g_s} we further assume:
\begin{subequations}
\begin{equation}
\summ{k} \sqrt{\lambda_k} \alpha_k^{r + \gamma - \frac{3}{2}} < \infty
\label{eqn:conds_g_b_4}
\end{equation}
\begin{equation}
\summ{k} \sqrt{\lambda_k}\alpha_k^{r + \delta - \frac{3}{2}} < \infty
\label{eqn:conds_g_s_22}
\end{equation}
\begin{equation}
\summ{k} \lambda_k \alpha_k^{2r + \beta + \gamma - 2} < \infty
\label{eqn:conds_g_b_2}
\end{equation}
\label{eqn:conds_g_b}
\end{subequations}
\begin{remark}
\label{rem:ell_2}
Consider the case where $A$ is a uniformly elliptic operator. From
\eqref{e:einf_ell} it is easy to see that conditions \eqref{eqn:conds_g_s} and
\eqref{eqn:conds_g_b} become
$$
\summ{k} \sqrt{\lambda_k}\alpha_k^{(r + d/4 -5/8 -\rho)} < \infty,
$$
and
$$
\summ{k } \lambda_k \alpha_k^{2r + d/2 - 17/8} < \infty,
$$
respectively.  
\end{remark}
We fix now an integer $p \geq 1$. We assume that the initial conditions are random
variables, independent of the $\sigma$--algebra generated by $W(t)$, with
\begin{equation}
\E \|x_0 \|^{2p} < \infty, \qquad \E \| y_0 \|^{2p} < \infty,
\label{e:ic_g_s}
\end{equation}
for $\gamma \in (0,2)$ and
\begin{equation}
\E \|x_0 \|^{2p} < \infty, \qquad \E \| y_0 \|^{4p} < \infty,
\label{e:ic_g_b}
\end{equation}
for $\gamma \in [2, \infty)$.

Now we state the convergence theorems. We start with $\gamma \in (0, 2)$.
\begin{theorem}
Let $x(t)$ be the solution of equation \eqref{e:motion_1} for $\gamma \in (0,2)$.
Assume that conditions \eqref{e:eigs_pos}, \eqref{eqn:conds_efs_g_s},
\eqref{e:b_cond}, \eqref{cond_coeff_h},\eqref{eqn:conds_lip}, \eqref{eqn:conds_g_s}
with $\rho = \frac{1}{2}$ and \eqref{e:ic_g_s} hold. Assume further that the initial
conditions for \eqref{e:ou_2} are stationary. Then $x(t)$ converges, as $\epsilon
\rightarrow 0$, to $X(t)$ which satisfies the following equation:
\begin{equation}
X(t) = x_0 + \int_0^t b(X(s)) \, ds + \int_0^t f(X(s))A^{-1} \, d W(s),
\label{limit_ito}
\end{equation}
the convergence being in  $L^{2p} (\Omega, C([0,T],\mathbb{R}))$:
\begin{equation}
\nrm{X(t) - x(t)}{p} \leq C \, \left( \epsilon^{ \gamma p} + \epsilon^{(2 -
\gamma)p - \sigma}  \right),
\label{etimate_thm_gamma_big}
\end{equation}
where $\sigma >0$ is arbitrarily small. The constant $C$ depends on the moments of the
initial conditions, the spectrum of the Wiener process, the operator $A$, the
exponent $p$, the maximum time $T$ and $\sigma$.
\label{thm:gamma_small}
\end{theorem}
In order to present the convergence theorems for the case $\gamma \geq 2 $ we need
to introduce some notation. We denote by $\Theta , \, \widehat{\Theta} : \ell_2 \rightarrow
\ell_2$ the diagonal operators defined by
\begin{equation}
\Theta = \mbox{diag} \left\{ \frac{\lambda_j}{2 \alpha_j^2} \right\}, \qquad
\widehat{\Theta} =\mbox{diag} \left\{ \frac{\lambda_j}{2 \alpha_j^2 (1 + \alpha_j)}
\right\}.
\label{e:chi}
\end{equation}
We will use the notation $\nabla \cdot A$ to denote the divergence of a matrix $A$,
i.e. $\{\nabla \cdot A \}_i = \sum_{j=1}^d A_{ij,j}$.

The next theorem covers the case $\gamma \in (2, \infty)$.
\begin{theorem}
Let $x(t)$ be the solution of equation \eqref{e:motion_1} for $\gamma \in (2,
\infty)$. Assume that conditions \eqref{e:eigs_pos}, \eqref{eqn:conds_efs_g_b},
\eqref{e:b_cond}, \eqref{cond_coeff_h},\eqref{eqn:conds_lip},
\eqref{eqn:conds_g_s_1} with $\rho = \frac{1}{2}$, \eqref{eqn:conds_g_s_2} with
$\rho = 0$, \eqref{eqn:conds_g_b} and  \eqref{e:ic_g_b} hold. Assume further that
the initial conditions for \eqref{e:ou_2} are stationary. Then $x(t)$ converges, as
$\epsilon \rightarrow 0$, to $X(t)$ which satisfies the following equation:
\begin{eqnarray}
X(t) & = & x_0 + \int_0^t b(X(s)) \, ds  +  \int_0^t  \nabla \cdot \left( f(X(s))
\Theta f^T(X(s)) \right) \, ds \nonumber \\ && - \int_0^t   f(X(s)) \Theta \nabla
\cdot f^T(X(s))  \, ds + \int_0^t f(X(s))A^{-1} \, d W(s) ,
\label{limit_stratonovich_1}
\end{eqnarray}
the convergence being in  $L^{2p} (\Omega, C([0,T],\mathbb{R}))$:
\begin{equation}
\nrm{X(t) - x(t)}{p} \leq C \left( \epsilon^{2p - \sigma} + \epsilon^{2p(\gamma
-2)-\sigma} \right),
%
%
\end{equation}
where $\sigma > 0$ is arbitrarily small. The constant $C$ depends on the moments of
the initial conditions, the spectrum of the Wiener process, the operator $A$
, the exponent $p$, the maximum time $T$ and $\sigma$.
\label{thm:gamma_big}
\end{theorem}
Finally, the case $\gamma =2$ is covered by the following theorem.
\begin{theorem}
Let $x(t)$ be the solution of equation \eqref{e:motion_1} for $\gamma = 2$.
 Assume that conditions \eqref{e:eigs_pos}, \eqref{eqn:conds_efs_g_b},
\eqref{e:b_cond}, \eqref{cond_coeff_h},\eqref{eqn:conds_lip},
\eqref{eqn:conds_g_s_1} with $\rho = \frac{1}{2}$, \eqref{eqn:conds_g_s_2} with
$\rho = 0$, \eqref{eqn:conds_g_b} and \eqref{e:ic_g_b} hold. Assume further that the
initial conditions for \eqref{e:ou_2} are stationary. Then $x(t)$ converges, as
$\epsilon \rightarrow 0$, to $X(t)$ which satisfies the following equation:
\begin{eqnarray}
X(t) & = & x_0 + \int_0^t b(X(s)) \, ds  +  \int_0^t \nabla \cdot \left( f(X(s))
\widehat{\Theta} f^T(X(s)) \right) \, ds  \nonumber \\ && - \int_0^t   f(X(s))
\widehat{\Theta} \nabla \cdot f^T(X(s)) \, ds + \int_0^t f(X(s))A^{-1} \, d W(s) .
\label{limit_stratonovich_3}
\end{eqnarray}
the convergence being in  $L^{2p} (\Omega, C([0,T],\mathbb{R}))$:
\begin{equation}
\nrm{X(t) - x(t)}{p} \leq C \, \epsilon^{2p -\sigma},
\label{etimate_thm_gamma_two}
\end{equation}
where $\sigma > 0$ is arbitrarily small. The constant $C$ depends on the moments of
the initial conditions, the spectrum of the Wiener process, the operator $A$
, the exponent $p$, the maximum time $T$ and $\sigma$.
\label{thm:gamma_two}
\end{theorem}
\begin{remark}
The second and third integrals in \eqref{limit_stratonovich_1} give the $d$--dimensional analogue
of the Stratonovich correction $\frac{1}{2} f(X) f'(X)$ in \eqref{e:lim2_intro}, when the system is
driven by an  infinite dimensional noise process. Similarly, the second and third integrals in
\eqref{limit_stratonovich_3}  correspond to the drift correction in equation
 \eqref{e:lim3_intro}.
\end{remark}
\begin{remark}
The assumptions of the convergence theorems ensure Lipschitz continuity and linear growth of
 all terms that appear in the limiting equations and, hence, existence and uniqueness of
solutions.  
\end{remark}
\begin{remark}
Throughout the paper we have set $\tau_0 = 1$, in order to simplify the notation. Of
course, the above convergence theorems hold true for arbitrary $\tau_0 >0$. In this
case, the matrix $widehat{\theta}$ defined in \eqref{e:chi} has to be modified:
\begin{equation}
\widehat{\Theta} =\mbox{diag} \left\{ \frac{\lambda_j}{2 \alpha_j^2 (1 + \tau_0
\alpha_j)} \right\}.
\label{e:theta_modif}
\end{equation}
Notice that we can formally retrieve the limiting equation for $\gamma < 2$ and
$\gamma > 2$ by sending $\tau_0$ in \eqref{e:theta_modif} to $\infty$ and $0$,
respectively.
\end{remark}
\begin{remark}
Problem \eqref{e:main} for $\gamma =0$ was considered in \cite{paper1_stuart}. It
was shown there that, under appropriate conditions on the spectrum of the Wiener
process and the operator $A$, the particle position $x(t)$ converges pathwise to the
solution $X$ of a second order SDE which we formally write as
$$
\ddot{X} = b(X) - \dot{X} + f(X)A^{-1} \dot{W}.
$$
It was proved in \cite{paper1_stuart} that the convergence rate is of
$\mathcal{O}(\eps^{2 - \sigma})$, where $\sigma >0$ is arbitrarily small. It is
natural, therefore, that the convergence rate in Theorem \ref{thm:gamma_small}
degenerates as $\gamma$ tends to either $0$ or $2$, since the limiting equation is
different in both cases.
\end{remark}
\subsection{Remarks on the Convergence Theorems}
\label{subsect:intro_remarks}
We present now a few comments on the convergence theorems. First, we note that the
smoothness assumptions on the eigenfunctions $\{\phi_k \}_{k =1}^{\infty}$
are more severe for $\gamma \in [2, \infty)$ than for $\gamma \in (0,2)$. This is
because, in order to prove our convergence theorems for $\gamma \geq 2$, we need
additional integrations by parts, using the It\^{o} formula. As a result, we need to
assume that more moments of the particle velocity at time $t =0$ exist when $\gamma
\geq 2$. Notice also that the convergence to the limiting equations becomes
arbitrarily slow as $\gamma \rightarrow 0$ and $\gamma \rightarrow 2^-$ in Theorem
\ref{thm:gamma_small}, as well as $\gamma \rightarrow 2^+$ in Theorem
\ref{thm:gamma_big}. This is also not surprising since the form of the coefficients in the limiting
equation is discontinuous at
$\gamma = 2$. The extensive numerical experiments reported in
\cite{paper3_stuart} indicate that the convergence rates of our Theorems are sharp.
On the other hand, the conditions that we have to impose on the spectrum of the
Wiener process, conditions (\ref{eqn:conds_g_s}) and (\ref{eqn:conds_g_b}), are not
sharp and are not--in general--independent from one another. In order to optimize
these conditions one needs more detailed information on the specific problem under
investigation, in particular on the properties of the eigenfunctions of the operator
$A$. Consider for example the case where $A$ is a uniformly elliptic operator,
Remarks \ref{rem:ell1} and \ref{rem:ell_2}.

Let us now try give an intuitive explanation of our results. First, for $\gamma < 2$ the
particle relaxation time
---which is of $\mathcal{O}(\epsilon^{\gamma})$--- is large compared
to the relaxation time of the noise--which is of $\mathcal{O}(\eps^2)$-- and consequently the
particles experience a rough noise with practically zero correlation time. This means that for
$\gamma < 2$ the OU process is not viewed from the point of view of the particle as a smooth
approximation to white noise and, as a result, the stochastic integral in the limiting equation
has to be interpreted in the It\^{o} sense. On the other hand, when $\gamma > 2$, the particle
relaxation time is small compared to that of the noise. Consequently, in this parameter regime
the rescaled OU process is indeed a smooth Gaussian approximation to white noise and the
stochastic integral in the limiting SDE should be interpreted in the Stratonovich sense as in
equation \eqref{limit_stratonovich_1}, in agreement with standard theorems \cite[sec. 10.3]
{arnold_l}. The case $\gamma \rightarrow \infty$ leads to the case of tracer
particles whose relaxation time is zero and covered precisely by these standard
theorems.

For the case $\gamma = 2$ the particle relaxation time is comparable in magnitude to
the noise correlation time and a resonance mechanism prevails which results in the limiting
 stochastic integral being neither that of  It\^{o} nor that of
Stratonovich. In this case the drift correction to the It\^{o} stochastic integral
depends on the detailed properties of the OU process, in particular its covariance.

It is well known that for second order stochastic differential equations the It\^{o}
and Stratonovich interpretations of the stochastic integral coincide. For certain Gaussian fields
$v(x,t)$  this also happens for the limiting equations given in our
convergence theorems: the Stratonovich correction, as well as its modified version
from Theorem \ref{thm:gamma_two}, will, in some situations, be identically zero due
to the specific properties of $v(x,t)$. In this case the limiting
equations are the same for all values of $\gamma$. This situation occurs for example
in the inertial particles problem which is discussed in section
\ref{sec:applications}, due to the fact that the fluid velocity is assumed to be
homogeneous and incompressible.

Let us now outline the method that we will use in order to prove the results of this
paper. The first step is to use the variation of constants formula to write the
particle velocity $y(t) :=\dot{x}(t)$ and particle position $x(t)$ as follows:
\begin{equation}
y(t) = y_0 \, e^{-\frac{t}{\epsilon^{\gamma}}} + \epsilon^{-\gamma} \int_0^t
e^{\frac{s - t}{\epsilon^{\gamma}}} \frac{v(x(s), s)}{\epsilon} \, ds +
\epsilon^{-\gamma} \int_0^t e^{\frac{s - t}{\epsilon^{\gamma}}} b(x(s)) \, ds
\label{y_soln_intro}
\end{equation}
and
\begin{eqnarray}
x(t)                  &   =  &   x_0 + \epsilon^{\gamma} y_0 (1
                                 - e^{-\frac{t} {\epsilon^{\gamma}}})
                                 +  \int_0^t
                 \frac{v(x(s),s)}
                                 {\epsilon} \, ds  + \int_0^t     b(x(s)) \, ds,
                      \nonumber \\ && -
                 \int_0^t e^{\frac{s - t}{\epsilon^{\gamma}}}  \frac{v(x(s),s)}
                                 {\epsilon} \, ds  -  \int_0^t e^{\frac{s -
                                 t}{\epsilon^{\gamma}}}  b(x(s)) \, ds,
\label{x_eqn_intro}
\end{eqnarray}
respectively. The next step is to use equations \eqref{y_soln_intro} and
\eqref{x_eqn_intro} in order to obtain sharp estimates on the moments of the
particle velocity. The basic strategy will be to derive first estimates valid for
$\gamma \in (0, \infty)$ and then use them in order to obtain sharper estimates
valid for $\gamma \in (0,2)$. We emphasize that sharper estimates for $\gamma \in
(0,2)$ are necessary for the proofs of the convergence theorems. Now, with the
estimates for the moments at hand we prove that the last two integrals on the right
hand side of equation \eqref{x_eqn_intro} are small in $L^{2p}(\Omega; C([0,T],
\mathbb{R}))$ for all values of $\gamma >0$.

Then we study the term which induces noise in (\ref{x_eqn_intro}), in the
limit $\epsilon \to 0$, namely the first integral on the
right hand side of this equation. We refer to this as $I(t)$. We use
the It\^{o} formula, together with the estimates on the moments of the particle velocity, to
show that $I(t)$ consists of an $\mathcal{O}(1)$ term plus higher order corrections.
The leading order  term in $I(t)$ is different for $\gamma < 2, \, \gamma  > 2$ and
$\gamma =2$: this is the term which is responsible for the difference in the
limiting equations for different $\gamma$. Finally, the proof of the convergence
theorems is completed by an application of Gronwall's lemma.

Throughout the paper we will make extensive use of estimates on the infinite
dimensional OU process $v(x,t)$ as well as the stochastic convolution $$\int_0^t
e^{\frac{s -t}{\epsilon^{\gamma}}} f(x(s)) A^{-1} \, d W(s),$$ Lemmas
\ref{lem:est_vel} and \ref{lemma:stoch_conv}, respectively. The proof of the first
of the above lemmas is based on Borell's inequality from the theory of Gaussian
processes \cite{adler}, while the proof of the second uses the factorization method
\cite{prato92}.

We remark that, unlike the methods used in the proofs of the convergence theorems in
\cite{paper1_stuart}, the proof in this paper relies on the presence of the friction
term $-\dot{x}$ in the equations of motion \eqref{eqn_motion_case_3}. The linear friction term
enables us to obtain representations \eqref{y_soln_intro} and \eqref{x_eqn_intro} for the
particle velocity and position, respectively, which are necessary for analyzing the dependence
of various moment bounds on $\epsilon$.
%
%
%
%
%
%
\section{Preliminary Results}
\label{sec:prelim_lemmas}
%
%
%
%
\subsection{The Integral Formulation}
\label{subsec:prelim_lemmas_int_form}
The first step is to obtain an integral equation for $x(t)$ that will be more
convenient for our analysis.
\begin{lemma}
Consider the equations of motion \eqref{e:motion_1}. Then the particle
position satisfies the following integral equation:
\begin{eqnarray}
x(t)                  &   =  &   x_0 + \epsilon^{\gamma} y_0 (1
                                 - e^{-\frac{t} {\epsilon^{\gamma}}})
                                 +  \int_0^t \left(1 - e^{\frac{s -
                                 t}{\epsilon^{\gamma}}}  \right)
                 \frac{v(x(s),s)}
                                 {\epsilon} \, ds
                \nonumber \\ &&  +
                 \int_0^t \left(1 - e^{\frac{s -
                                 t}{\epsilon^{\gamma}}}  \right) b(x(s)) \, ds.
\label{x_eqn}
\end{eqnarray}
\end{lemma}
\begin{proof} We start by solving the equation for $y(t) = \dot{x}(t)$ using the variation
of constants formula:
\begin{equation}
y(t) = y_0 \, e^{-\frac{t}{\epsilon^{\gamma}}} + \epsilon^{-\gamma} \int_0^t
e^{\frac{s - t}{\epsilon^{\gamma}}} \frac{v(x(s), s)}{\epsilon} \, ds +
\epsilon^{-\gamma} \int_0^t e^{\frac{s - t}{\epsilon^{\gamma}}} b(x(s)) \, ds.
\label{y_soln}
\end{equation}
Another integration will give us an integral equation for $x(t)$ which involves a
double integral:
\begin{eqnarray}
x(t)         & =  &    x_0 + \int_0^t y_0 \,
                       e^{-\frac{s}{\epsilon^{\gamma}}} \, ds  +
                       \frac{1}{\epsilon^{\gamma}} \int_0^t \left( \int_0^{\ell}
                       e^{\frac{s - \ell}{\epsilon^{\gamma}}}
                       H(x(s),s) \, ds \right) d \ell
           \nonumber \\ & = &
                        x_0 + \epsilon^{\gamma} y_0 (1 - e^{-\frac{t}
                        {\epsilon^{\gamma}}}) + I(t),
\label{xinterm}
\end{eqnarray}
where
\begin{equation}
H(s) = \frac{v(x(s),s)}{\epsilon} + b(x(s)).
\nonumber
\end{equation}
We can reduce $I(t)$ to a single integral as follows: First we define the following
function:
\begin{equation}
F(\ell) = \int_0^{\ell} e^{\frac{s} {\epsilon^{\gamma}}} H( s) \, ds.
\nonumber
\end{equation}
Now we perform an integration by parts:
\begin{eqnarray}
I(t)                  & = &  \frac{1}{\epsilon^{\gamma}} \int_0^t
                          \left( \int_0^{\ell}
                          e^{\frac{s - \ell}{\epsilon^{\gamma}}}
                          \left( \frac{v(x(s), s)}{\epsilon} +b(s) \right)
              \, ds \right) d \ell
                   \nonumber \\ & = &
                           \frac{1}{\epsilon^{\gamma}}
                           \int_0^t
                           e^{-\frac{\ell}{\epsilon^{\gamma}}}F(\ell)
                           \, d \ell
                   \nonumber \\ & = &
                            -  F(\ell) e^{-\frac{\ell}
                            {\epsilon^{\gamma}}} \big|^{\ell = t}_{\ell =
                            0} + \int_0^t e^{- \frac{\ell}
                            {\epsilon^{\gamma}}} \, dF(\ell)
                   \nonumber \\ & = &
                           -  e^{-\frac{t}{\epsilon^{\gamma}}}
                            \int_0^t
                          e^{\frac{s}{\epsilon^{\gamma}}}
                            H(s) \, ds + \int_0^t H(s) \, ds
                   \nonumber \\ & = &
                            \int_0^t \left( 1 -
                          e^{\frac{s - t}{\epsilon^{\gamma}}}
                          \right) \frac{v(x(s), s)}{\epsilon} \, ds +
              \int_0^t \left( 1 -
                          e^{\frac{s - t}{\epsilon^{\gamma}}}
                          \right) b(x(s)) \, ds.
\label{ixexpr}
\end{eqnarray}
Substituting (\ref{ixexpr}) into (\ref{xinterm}) we obtain (\ref{x_eqn}). \qquad\end{proof} 

\bigskip

Throughout the paper we will use the following notation:
\begin{subequations}
\begin{equation}
I_1(t)  = \epsilon^{\gamma} y_0 (1  - e^{-\frac{t}
       {\epsilon^{\gamma}}}),
\label{term_i_1}
\end{equation}
\begin{equation}
I_2(t)  =  \int_0^t \frac{f(x(s))\eta(s)}{\epsilon} \, ds,
\label{term_i_2}
\end{equation}
\begin{equation}
I_3(t)  =   - \int_0^t e^{\frac{s - t}{\epsilon^{\gamma}}}
           \frac{f(x(s))\eta(s)} {\epsilon} \, ds,
\label{term_i_3}
\end{equation}
\begin{equation}
I_4(t)  =   - \int_0^t e^{\frac{s - t}{\epsilon^{\gamma}}}
            b(x(s)) \, ds.
\label{term_i_4}
\end{equation}
\end{subequations}
Using this notation the particle position $x(t)$ can be written in the form
\begin{equation}
x(t) = x_0 + \sum_{i=1}^4 I_i(t) + \int_0^t b(x(s)) \, ds.
\label{e:x_ii}
\end{equation}
We clearly have:
\begin{equation}
\nrm{I_1(t)}{p} \leq C \, \epsilon^{2 \gamma p}.
\label{i1_b}
\end{equation}
Now we want to study terms $I_2(t), \, I_3(t)$ and $I_4(t)$. As explained in section
\ref{subsect:intro_remarks}, we want to show that $I_3(t)$ and $I_4(t)$ are $o(1)$
in $L^{2p} (\Omega, C([0,T],\mathbb{R}))$ for every $\gamma \in (0,\infty)$ and then
show that the behavior of the term $I_2(t)$ as $\epsilon \rightarrow 0 $ depends on
$\gamma$. In order to obtain the necessary bounds we will need sharp estimates on
the moments of the particle velocity. We will obtain these estimates in section
\ref{subsec:prelim_lemmas_y_moments}. Before doing this, we need some estimates on
the velocity field $v(x,t)$.
%
%
%
%
\subsection{Estimates on the Colored Noise}
\label{subsec:prelim_lemmas_col_noise}
In this subsection we present two results which will be used in the proofs of the
convergence theorems. We start with an estimate on the infinite dimensional OU
process.
\begin{lemma}
\label{lem:est_vel}
Assume that conditions \eqref{eqn:conds_efs_g_b}, \eqref{cond_coeff_h} and
\eqref{eqn:conds_g_s_1} with $\rho \in \mathbb{R}$ are satisfied. Then the following
estimate holds:
\begin{equation}
\nrm{A^{-\rho} v(x(t),t)}{p} \leq C \, \epsilon^{- \sigma},
\label{est_vel_gauss}
\end{equation}
where $\sigma > 0$ is arbitrarily small.
\end{lemma}
\begin{proof} Let $\eta(t) = \{\eta_k(t) \}_{k=1}^{\infty}:\ell_2 \rightarrow \R$ be
the solution of \eqref{eqn:ou_k} with stationary initial conditions. The $k$th
component $\eta_k(t)$ solves the equation
\begin{equation*}
d \eta_k = - \alpha_k \eta_k  \, dt +  \sqrt{\lambda_k} d W_k.
\end{equation*}
A simple variant of Theorem A.1 from \cite{paper1_stuart} yields:
\begin{equation}
\mathbb{E} \left( \sup_{0 \leq t \leq T} |\eta_k(t) |^{2 p} \right) \leq \left(
\frac{\lambda_k}{\alpha_k} \right)^{2p} \left( 1 + \ln \left( \frac{\alpha_k
T}{\epsilon^2} \right) \right).
\label{eqn:ou_estim_component}
\end{equation}
Let $Y(x(t), t)  :=   A^{-\rho} v(x(t),t)$. We first consider the case $p = 1$. We
have:

\begin{eqnarray}
\nrm{ Y(x(t),t)}{}    & = &
                 \normm{ \summ{k} h_k \phi_k(x(t))  \alpha_k^{- \rho} \eta_k(t)}
             \nonumber \\ & = &
                    \mathbb{E} \left( \sup_{0 \leq t \leq T} \left\| \summ{k, \ell} h_k \cdot
                    h_{\ell} \phi_k(x(t)) \phi_{\ell}
          (x(t)) \alpha_k^{- \rho} \alpha^{- \rho}_{\ell} \eta_k(t)
          \eta_{\ell}(t) \right\| \right)
             \nonumber \\ & \leq &
                  \mathbb{E} \left( \sup_{0 \leq t \leq T} \left( \sum_{k ^{+}} \|
                  h_k \|^2  |\phi_k(x(t))|^2\alpha^{-2 \rho}_k
                  |\eta_{k}(t)|^2 \right) \right)
               \nonumber \\  &&  +
                   \mathbb{E} \left( \sup_{0 \leq t \leq T}
                  \left( \summ{k} \sum_{\ell \neq k} \| h_k
                  \|  \| h_{\ell} \| \phi_k(x(t)) \phi_{\ell}(x(t))
                  \alpha_k^{- \rho} \alpha_{\ell}^{- \rho} \eta_k(t) \eta_{\ell}(t)
                  \right) \right)
               \nonumber \\ & \leq &
                  C \,  \summ{k} \frac{|k|^{2(r +
                  \alpha)}}{\alpha_k^{2 \rho}} \, \mathbb{E} \left(
                  \sup_{0 \leq t \leq T} | \eta_k(t) |^2 \right) +
                  C \,  \left( \summ{k} \frac{|k|^{r +
                  \alpha}}{\alpha_k^{\rho}} \, \mathbb{E} \left(
                  \sup_{0 \leq t \leq T} | \eta_k(t) | \right)
                  \right)^2
                \nonumber \\ & \leq &
                 C \, \left( \summ{k}\frac{ \lambda_k |k|^{2(r +
                 \alpha)}}{\alpha_k^{1 + 2 \rho}} \right)
                 \epsilon^{-\sigma} + C \, \left( \summ{k}
                 \frac{ \sqrt{\lambda_k} |k|^{r + \alpha}}
                 {\alpha_k^{\frac{1}{2} +  \rho}} \right)
                 \epsilon^{-\sigma}
               \nonumber \\ & \leq &
                 C \, \epsilon^{-\sigma},
\nonumber
\end{eqnarray}
on account of condition \eqref{eqn:conds_g_s_1}. We can proceed in the same way for
$p > 1$, by breaking the sums into various parts, until we have sums that involve
independent OU processes. Condition \eqref{eqn:conds_g_s_1} ensures the summability
of all the sums that appears. The lemma is proved.
\qquad\end{proof} 

\bigskip

Using the above lemma we can easily obtain the following estimate.
\begin{lemma}
\label{lem:est_vel_int}
Assume that the conditions of Lemma \ref{lem:est_vel} hold. Define
\begin{equation}
I(t) =  \int_0^t e^{\frac{s - t}{\epsilon^{\gamma}}}  A^{- \rho} v(x(s),s) \, ds.
\label{e:v_in}
\end{equation}
Then the following estimate holds:
\begin{equation}
\nrm{I(t)}{p}  \leq C \, \epsilon^{ 2p \gamma - \sigma},
\label{e:vel_int}
\end{equation}
where $\sigma > 0$ is arbitrarily small.
\end{lemma}
\begin{proof} We have, for $t \in [0, T]$:
\begin{eqnarray}
\|  I(t) \|^{2p} & = &  \left| \left|    \int_0^t
                           e^{\frac{s - t}{\epsilon^{\gamma}}} A^{-\rho} v(x(s),s)
               \, ds \right| \right|^{2p}
                          \nonumber \\ & \leq &
                           \sup_{0 \leq s \leq T} \| A^{-\rho} v(x(s),s) \|^{2p}
              \left( \int_0^t e^{\frac{s-t}{\epsilon^{\gamma}}}
              \, ds  \right)^{2p}
                          \nonumber \\ & \leq &
                          \epsilon^{2p \gamma}
               \sup_{0 \leq t \leq T}
              \| A^{-\rho} v(x(s),s) \|^{2p}.
\nonumber
\end{eqnarray}
Lemma \ref{lem:est_vel} now yields estimate \eqref{e:vel_int}. \qquad\end{proof} 
\begin{remark}
\label{rem:vel}
The techniques used in the proof of Lemma \ref{lem:est_vel} enable us to conclude that we can
bound uniformly all moments of the field $v(x,t)$:
\begin{equation}
\E \|A^{- \rho} v(x,t) \|^{2p} \leq C,
\nonumber
\end{equation}
provided that the assumptions of the Lemma are satisfied. Furthermore, the method of
proof of Lemma \ref{lem:est_vel_int} gives:
\begin{equation}
\mathbb{E} \|I(t) \|^{2p} \leq C \, \epsilon^{ 2p \gamma},
\nonumber
\end{equation}
where $I(t)$ is defined in \eqref{e:v_in}.  
\end{remark}
\begin{remark}
Lemma \ref{lem:est_vel_int} with $\rho = 0$ provides us with estimates for
$I_3(t)$ and $I_4(t)$ which we will use for $\gamma \in [2, \infty)$:
\begin{subequations}
\begin{equation}
\nrm{I_3(t)}{p}         \leq  C \, \epsilon^{2p( \gamma -1) - \sigma}
                          , \; \; \; \gamma \in  [2, \infty ),
\label{i3_g_b}
\end{equation}
and
\begin{equation}
\nrm{I_2(t)}{p}         \leq  C \, \epsilon^{-2p - \sigma}
                          , \; \; \; \gamma \in  [2, \infty ),
\label{i2_g_b}
\end{equation}
\label{e:i23_g_b}
\end{subequations}
where $\sigma >0$ is arbitrarily small. This estimate is not sharp enough when
$\gamma \in (0,2)$ and we need to improve it. This will be accomplished in Corollary
\ref{cor:i3_bd_g_s}. 
\end{remark}
\begin{remark}
Assume that the moments of the particle velocity $y(t)$ satisfy
$$
\nrm{y(t)}{p} \leq C \epsilon^{\zeta p},
$$
for some $\zeta \in \R$. A repeated use of H\"{o}lder's inequality,
together with the Gaussianity of the process $\eta(t)$ as in the proof of Lemma 4.3
in \cite{paper1_stuart} enables us to prove that
\begin{equation}
\E \left( \sup_{0 \leq t \leq T} \|y(t) \|^{2p} \| \eta(t)\|^{2 n}_{\ell^2} \right) \leq
\epsilon^{- \zeta p-\sigma},
\label{e:bootstrap}
\end{equation}
for every $n$, assuming that $\mbox{Tr}(Q) < \infty$, with $\sigma >0$, arbitrarily
small. In the sequel we will have the occasion to use estimate
\eqref{e:bootstrap} and variants of it repeatedly. 
\end{remark}
We proceed now with an estimate on a stochastic integral.
\begin{lemma}
\label{lemma:stoch_conv}
Consider the stochastic integral
\begin{equation}
I(t) = \int_0^t e^{\frac{s - t}{\epsilon^{\gamma}}} f(x(s))A^{-1} \, dW(s).
\nonumber
\end{equation}
Assume that conditions \eqref{eqn:conds_efs_g_s}, \eqref{cond_coeff_h} and
\eqref{eqn:conds_g_s_1} with $\rho = \frac{1}{2}$ hold. Then we have the following
estimate:
\begin{equation}
\nrm{ I(t)}{p}  \leq C \, \epsilon^{p\gamma - \sigma}.
\label{e:stoch_int_est}
\end{equation}
where $\sigma > 0$ is arbitrarily small.
\end{lemma}
\begin{proof}
 We fix $\alpha \in (0, \frac{1}{2})$ and use the factorization method from
\cite[sec. 5.3]{prato92} to obtain
\begin{eqnarray}
I(t) & := & \int_0^t e^{ \frac{s-t}{\epsilon^{\gamma}}} f(x(s))
       A^{-1}dW(s)
   \nonumber \\ & = &
        \frac{\sin(\pi \alpha)}{\pi} \int_0^t e^{ \frac{s-t}{\epsilon^{\gamma}}}
        (t - s)^{\alpha -1} Y(s) \, ds
\nonumber
\end{eqnarray}
where
\begin{equation}
Y(s) = \int_0^s e^{ \frac{\sigma -s}{\epsilon^{\gamma}}} (s - \sigma)^{-\alpha}
f(x(\sigma)) A^{-1} \, dW(\sigma).
\nonumber
\end{equation}
We choose $m > \frac{1}{2 \alpha}$ and use H\"{o}lder inequality to obtain:
\begin{equation}
\|I(t) \|^{2 m} \leq C \left( \int_0^t  \left| e^{\frac{s-t}
                    {\epsilon ^{\gamma}}} (t - s)^{\alpha -1}
                    \right|^{\frac{2m }{2m - 1}} ds \right)
                    ^{2m -1} \int_0^t
                    \|Y(s) \|^{2m} \, ds.
\nonumber
\end{equation}
A change of variables now yields:
\begin{eqnarray}
J(t) & := & \int_0^t  \left| e^{\frac{s-t}
                    {\epsilon ^{\gamma}}} (t - s)^{\alpha -1}
                    \right|^{\frac{2m }{2m - 1}} \, ds
       \nonumber \\ & = &
             \left( \frac{2m -1}{2m} \right)^{\frac{2m -1}{2m}(\alpha +
             2(m-1))} \epsilon^{\gamma \frac{2m \alpha -1}{2m -
             1}}\int^{t\frac{2m}{2m-1}\epsilon^{-\gamma}}_0 e^{-z}
             z^{\frac{2m}{2m-1}(\alpha -1)} \, dz
       \nonumber \\ & \leq &
             \epsilon^{\gamma \frac{2m \alpha -1}{2m -
             1}}\int^{\infty}_0 e^{-z}
             z^{\frac{2m}{2m-1}(\alpha -1)} \, dz
        \nonumber \\ & \leq &
              C \, \epsilon^{\gamma \frac{2m \alpha -1}{2m - 1}}.
\nonumber
\end{eqnarray}
In the above estimate we used the fact that, since $m > \frac{1} {2 \alpha}$, we
have $e^{-z} z^{\frac{2m}{2m-1}(\alpha -1)} \in L^1(\mathbb{R}^+)$. Consequently, we
have:
\begin{equation}
\nrm{I(t)}{m} \leq C \, \epsilon^{\gamma (2m \alpha -1)} \mathbb{E} \int^T_0 \|Y(s)
\|^{2m} \, ds.
\nonumber
\end{equation}
To proceed further, we use \cite[Lemma 7.2]{prato92} to deduce that there exists a
constant $C_m >0$ depending only on $m$ such that
\begin{equation}
\sup_{0 \leq s \leq T} \mathbb{E} \|Y(s) \|^{2m} \leq C_m \mathbb{E} \left(
\int_0^{s} e^{-2 \frac{s - \sigma}{\epsilon^{\gamma}}} (s - \sigma)^{-2 \alpha}
\|f(x(\sigma)) A^{-1} \|^2_{L^0_2} \, d \sigma \right)^m,
\nonumber
\end{equation}
with
\begin{eqnarray}
\|f(x(\sigma)) A^{-1} \|^2_{L^0_2} &:=& \mbox{Tr} \left[ \left(
                                f(x(\sigma)) A^{-1} \right)Q \left(f(x(\sigma))
                                A^{-1} \right)^* \right]
      \nonumber \\ & = &
                 \summ{k} \frac{\lambda_k |\phi_k(x(\sigma))|^2 \|h_k \|^2}
                 {\alpha_k^2}
         \nonumber \\ & \leq &
               C  \summ{k} \lambda_k |\alpha_k|^{2(\alpha + r -1)} < \infty,
\label{e:trace}
\end{eqnarray}
on  account of condition \eqref{eqn:conds_g_s_1} with $\rho = \frac{1}{2}$. Now we can
apply the same change of variables that we used in the estimate for $J(t)$ to obtain:
\begin{eqnarray}
\sup_{0 \leq s \leq T} \mathbb{E} \|Y(s) \|^{2m}
                     & \leq &  C \, \mathbb{E} \left( \int_0^{s} e^{2
                     \frac{ \sigma - s}{\epsilon^{\gamma}}} (s -
                     \sigma)^{-2 \alpha} \, d \sigma \right)^m
               \nonumber \\  & \leq &
                      C \, \left(\epsilon^{\gamma(1 - 2 \alpha )} \int_0^{\frac{2s}
                      {\epsilon^{\gamma}} } e^{-z} z^{-2 \alpha} \, dz  \right)^{m}
                \nonumber \\ & \leq &
                       C \, \epsilon^{\gamma m (1 - 2 \alpha )}.
\nonumber
\end{eqnarray}
From the above estimates we conclude:
\begin{equation}
\nrm{I(t)}{ m} \leq C \, \epsilon^{\gamma (m-1)}.
\nonumber
\end{equation}
Now estimate \eqref{e:stoch_int_est} follows for $p > \frac{1}{2 \alpha}$ upon
taking $p = m$. For $p \leq \frac{1}{2 \alpha}$ we apply H\"{o}lder inequality to
obtain:
\begin{equation}
\nrm{I(t)}{p} \leq \left( \nrm{I(t)}{ m} \right)^{\frac{p}{m}} \leq C \,
\epsilon^{\gamma p - \frac{p}{m}},
\nonumber
\end{equation}
which completes the proof of the lemma, since $m$ can be chosen to be arbitrarily
large. \qquad\end{proof} 
%
%
%
%
\subsection{Bounds on the Moments of $y(t)$}
\label{subsec:prelim_lemmas_y_moments}
In this subsection we will obtain bounds on the moments of $y(t)$ that we will need
for the convergence theorem. In order to obtain estimates on the moments of the
particle velocity we first need to obtain a crude estimate on the moments of the
particle position. This estimate will be improved later.
\begin{lemma}
Let $x(t)$ satisfy equation (\ref{e:motion_1}). Assume that \eqref{e:b_cond} and
\eqref{e:ic_g_s} as well as the conditions of Lemma \ref{lem:est_vel} with $\rho =
0$ are satisfied. Then the following estimate holds:
\begin{equation}
\nrm{x(t)}{p} \leq C \, \epsilon^{-2p - \sigma} ,
\label{e:mom_x_1}
\end{equation}
where $\sigma > 0$ is arbitrarily small.
\label{lem:mom_x_1}
\end{lemma}
\begin{proof} The particle position is given by \eqref{x_eqn}, which can be written in the
form
$$
x(t) = x_0 + I_1(t) + I_2(t) + I_3(t) + \int_0^t \left(1 -
e^{\frac{t-s}{\eps^{\gamma}}} \right) b(x(s)) \, ds.
$$
The Lipschitz continuity, assumption \eqref{e:b_cond}, of $b(x)$ implies that there
exists $C >0$ such that
\begin{equation}
\|b(x) \| \leq C (1 + \|x \|).
\label{e:b_est}
\end{equation}
We use this, together with estimates \eqref{i1_b} and \eqref{e:i23_g_b} as well as
Lemma \ref{lem:est_vel} to obtain
\begin{eqnarray}
\nrm{x(t)}{p}  & \leq &  \E \|x_0 \|^{2p} + C \epsilon^{ 2 \gamma p }  + C \eps^{-2p
                          -\sigma}  +      C \epsilon^{ 2p (\gamma -1) - \sigma }
               \nonumber \\ &&
                          + C \left(1 + \int_0^T \nrm{x(t)}{p} \, dt \right)
              \nonumber \\ & \leq &
                  C \eps^{-2p  -\sigma} +   \int_0^T \nrm{x(t)}{p} \, dt.
\nonumber
\end{eqnarray}
Estimate \eqref{e:mom_x_1} now follows from Gronwall's lemma. \qquad\end{proof} 

\bigskip

We have already mentioned that the bounds on the moments of the particle velocity will be
different for $\gamma < 2$ and $\gamma \geq 2$. We start with the regime
$\gamma \in [2, \infty)$.
\begin{lemma}
Let $x(t)$ satisfy equation (\ref{e:motion_1}) an let  $y(t) = \dot{x}(t)$.
Assume that assumptions \eqref{e:b_cond} and \eqref{e:ic_g_s}
as well as the conditions of Lemma \ref{lem:est_vel} with $\rho = 0$ are satisfied.
Then the following estimate holds:
\begin{equation}
\nrm{y(t)}{p} \leq C \, \epsilon^{-2p - \sigma} ,
\label{est:lem:mom_g_b}
\end{equation}
where $\sigma > 0$ is arbitrarily small.
\label{lem:mom_g_b}
\end{lemma}
\begin{proof} The particle velocity is given by \eqref{y_soln}, which can be written in the
form
$$
y(t) = y_0 e^{-\frac{t}{\epsilon^{\gamma}}} - \epsilon^{-\gamma} I_3(t) -
\epsilon^{-\gamma} I_4(t).
$$
Lemma \ref{lem:mom_x_1} and  estimate \eqref{e:b_est} give
$$
\nrm{I_4(t)}{p} \leq C \eps^{2p(\gamma -1) -\sigma}.
$$
We use the above estimate, together with \eqref{i3_g_b} to obtain
\begin{eqnarray}
\nrm{y(t)}{p}  & \leq &  C_1 + C_2 \epsilon^{ -2 \gamma p } \nrm{I_3(t)}{p} + C_3
                               \epsilon^{ -2 \gamma p } \nrm{I_4(t)}{p}
      \nonumber \\ & \leq &
                              C ( \epsilon^{-2 p - \sigma} + 1),
\nonumber
\end{eqnarray}
from which the estimate follows. \qquad\end{proof} 

\bigskip

Estimate (\ref{est:lem:mom_g_b}) will be sufficient for our purposes for $\gamma \in
[2, \infty)$. However, it is not sharp enough for $\gamma \in (0,2)$. In order to
prove the convergence theorem for values of $\gamma$ in this parameter regime  we
need to improve the estimate for the $2p$th moments of the particle position and
particle velocity. For these two estimates we need some preliminary estimates which
will be also used in the proof of the convergence theorem. We start with the
following lemma.
\begin{lemma}
\label{lem:y_int}
Assume that conditions \eqref{e:eigs_pos}, \eqref{eqn:conds_efs_g_s}, \eqref{cond_coeff_h},
together with \eqref{eqn:conds_g_s_2} with $\rho \in \R$ hold. Fix $x,y
\in \R^d$ and define $df(x)y \in L (\ell_2, \R^d)$ by
$$
\left\{(df)y \right\} \gamma = \summ{k} h_k y \cdot \nabla \phi_k  \gamma_k.
$$
Let
$$I(t) = \int_0^t e^{\frac{s - t}{\epsilon^{\gamma}}} df(x(s))y(s)
      A^{-\rho} \eta(s) \, ds
$$
and
\begin{equation}
\widehat{I}(t) = \int_0^t df(x(s)) y(s) A^{-\rho} \eta(s) \, ds.
\label{e:int_hat}
\end{equation}
Then the following estimates hold:
\begin{equation}
\nrm{I(t)}{p} \leq C \epsilon^{(2p-1)\gamma - 2p - \sigma},
\label{e:y_in_est}
\end{equation}
and
\begin{equation}
\nrm{\widehat{I}(t)}{p} \leq C \epsilon^{-2p - \sigma},
\label{e:y_df_est}
\end{equation}
where $\sigma >0$ is arbitrarily small.
\end{lemma}
\begin{proof} An application of H\"{o}lder inequality yields:
\begin{eqnarray}
 \| I(t) \|^{2p}   & \leq & \left( \int_0^t e^{\left( \frac{s - t}{\epsilon^{\gamma}}
 \frac{2p}{2p-1} \right)}  \, ds  \right)^{2p -1}
              \int_0^t \| df(x(s))y(s) A^{-\rho} \eta(s) \|^{2p} \, ds
         \nonumber \\ & \leq &
 \epsilon^{(2p-1) \gamma }
              \int_0^t \| df(x(s))y(s) A^{-\rho} \eta(s) \|^{2p} \, ds.
\nonumber
\end{eqnarray}
A simple variant of \eqref{e:bootstrap}, together with calculations similar to
those used in the proof of Lemma \ref{lem:est_vel}, gives
\begin{eqnarray}
\nrm{I(t)}{p} & \leq & \epsilon^{(2p-1)\gamma} \int_0^T  \E \|df(x(s)) y(s)
 A^{- \rho} \eta(s) \|^{2p} \, ds
         \nonumber \\   & \leq &
C \epsilon^{(2p-1)\gamma} \int_0^T  \E \left( \summ{i} \left| \summ{j} \sum_{k = 1}^d
f_{ij,k}(x(s)) y_k(s) \alpha_j^{-\rho} \eta_j(s) \right|^{2} \right)^p \, ds
       \nonumber \\ & \leq &
C \epsilon^{(2p-1)\gamma} \int_0^T  \E \left( \left( \summ{j} \alpha_k^
{\beta + r - \rho} \eta_j(s) \right)^{2p} \| y(s)  \|^{2p} \right)
        \nonumber \\ & \leq &
 C \epsilon^{(2p-1)\gamma - 2p - \sigma},
\nonumber
\end{eqnarray}
assuming that condition \eqref{eqn:conds_g_s_2} holds. This proves
\eqref{e:y_in_est}. The proof of \eqref{e:y_df_est} is almost identical and is
omitted.
\qquad\end{proof} 

\bigskip

We introduce some notation that we will use repeatedly throughout the rest of the
paper. We set:
\begin{eqnarray}
J_1(t) & := & \left[ A^{-1} v(x(t),t) - e^{-\frac{t}{\epsilon^{\gamma}}} v(x_0,0)
\right], \; \; J_2(t) :=  \int_0^t e^{\frac{s - t}{\epsilon^{\gamma}}} df(x(s)) y(s)
A^{- 1} \eta(s) \, ds,
\nonumber \\ J_3(t) & := & \int_0^t e^{\frac{s - t}{\epsilon^{\gamma}}}  f(x(s))
A^{- 1} \, dW(s) , \; \; \; \; \; \; \; \; J_4(t) :=  \int_0^t e^{\frac{s -
t}{\epsilon^{\gamma}}} A^{- 1} v(x(s),s) \, ds.
\nonumber
\end{eqnarray}
Lemmas \ref{lem:est_vel}, \ref{lem:est_vel_int} \ref{lemma:stoch_conv} and
\ref{lem:y_int}, and assuming that the initial conditions for $v(x,t)$ are stationary and that
conditions \eqref{eqn:conds_g_s_1} with $\rho =\frac{1}{2}$ and \eqref{eqn:conds_g_s_2} with
$\rho = 1$ are satisfied, provide us with the following bounds:
\begin{eqnarray}
\nrm{J_1(t)}{p} & \leq & C \epsilon^{-\sigma}, \; \; \; \; \; \; \, \nrm{J_2(t)}{p}
\leq C \epsilon^{(2p-1) \gamma - 2p-\sigma},
\nonumber \\ \nrm{J_3(t)}{p} & \leq & C \epsilon^{p \gamma -\sigma} ,    \; \; \; \;
\nrm{J_4(t)}{p}  \leq  C \epsilon^{2 p \gamma-\sigma}.
\label{e:ji_est}
\end{eqnarray}
\begin{lemma}
\label{lem:mom_x_2}
Let $x(t)$ be given by \eqref{x_eqn} and let $y(t) = \dot{x}(t)$ be given by
\eqref{y_soln}. Suppose that conditions \eqref{e:b_cond} and \eqref{eqn:conds_g_s}
with $\rho =1$ hold. Then for $\gamma \in (0,2)$ the following estimate holds:
\begin{equation}
\nrm{x(t)}{p} \leq C.
\label{e:mom_x_2}
\end{equation}
\end{lemma}
\begin{proof} Consider first the term $I_2(t)$ defined in \eqref{term_i_2}. We integrate by
parts to obtain:
\begin{eqnarray}
I_2(t)     & = & \int_0^t f(x(s)) A^{-1} \, dW(s) - \epsilon \widehat{J}_1(t)
         + \epsilon \widehat{J}_2(t),
\label{e:i2_thm}
\end{eqnarray}
where
$$
\widehat{J}_1(t) := \left[ A^{-1} v(x(t),t) - A^{-1} v(x(0),0)  \right],
$$
and
$$
\widehat{J}_2(t) = \int_0^t df(x(s)) y(s) A^{-1} \eta(s) \, ds.
$$
Lemma \ref{lem:est_vel} with $\rho = 1$ gives
\begin{equation}
\nrm{\epsilon \widehat{J}_1(t)}{p} \leq C \, \epsilon^{2p - \sigma}.
\nonumber
\end{equation}
Moreover, Lemma \ref{lem:y_int} with $\rho =1$, in particular estimate
\eqref{e:y_df_est}, yields
\begin{equation}
\nrm{\epsilon \widehat{J}_2(t)}{p} \leq C.
\nonumber
\end{equation}
Furthermore, the Burkholder--Davis--Gundy inequality, assuming that condition
\eqref{eqn:conds_g_s_1} with $\rho = \frac{1}{2}$ holds, yields
$$
\nrm{\int_0^t f(x(s)) A^{-1} \, dW(s)}{p} \leq C.
$$
We put the above estimates together to conclude that
$$
\nrm{I_2(t)}{p} \leq C.
$$
We use this estimate, together with \eqref{i3_g_b} and the assumptions
\eqref{e:ic_g_s} in equation \eqref{x_eqn} to obtain:
\begin{eqnarray}
\nrm{x(t)}{p} & \leq  & C \E \|x_0 \|^{2p} + C \E \|y_0 \|^{2p} + C \eps^{2p(\gamma
                         -1)- \sigma} + C \int_0^T \nrm{x(t)} \, dt
              \nonumber \\ & \leq &
                     C + \int_0^T \nrm{x(t)} \, dt. \nonumber
\end{eqnarray}
Estimate \eqref{e:mom_x_2} now follows from Gronwall's lemma. \qquad\end{proof} 

\bigskip

Now we are ready to obtain a sharp bound on the moments of the particle velocity for
$\gamma \in (0,2)$.
\begin{lemma}
Let $x(t)$ be the solution of \eqref{e:motion_1} and let and $y(t) = \dot{x}(t)$ .
Assume that $\mathbb{E} \|y_0 \|^{2p} < \infty$ and that conditions
\eqref{eqn:conds_efs_g_s} ,\eqref{e:b_cond}, \eqref{cond_coeff_h} and
\eqref{eqn:conds_g_s_1} with $\rho = \frac{1}{2}$ and \eqref{eqn:conds_g_s_2} with
$\rho =1$ are satisfied. Then the following estimate holds:
\begin{equation}
\nrm{y(t)}{p} \leq C \epsilon^{- \gamma p - \sigma}, \; \; \gamma \in (0,2).
\label{eqn:mom_g_s}
\end{equation}
\label{lem:mom_g_s}
\end{lemma}
\begin{proof} Lemma \ref{lem:mom_x_2} and  estimate \eqref{e:b_est} give
\begin{equation}
\nrm{I_4(t)}{p} \leq C \eps^{2p\gamma }.
\label{i4_b}
\end{equation}
We combine this estimate with \eqref{y_soln} to deduce:
$$
\nrm{y(t)}{p} \leq C_1 + C_2 \epsilon^{-2 p \gamma} \nrm{I_3(t)}{p}.
$$
We need to get a sharper estimate on $I_3(t)$ than \eqref{i3_g_b}. For
this we need to integrate by parts. We apply It\^{o} formula to the function
$$G_{ik}(s,x,\eta) = e^{\frac{s}{\epsilon^{\gamma}}} f_{ik}(x(s))\frac{\eta_k(s)}{\alpha_k},$$
to obtain, after some algebra:
\begin{eqnarray}
I_3(t)       & = &
      \epsilon \left[  A^{-1} v(x(t),t) -
      e^{-\frac{t}{\epsilon^{\gamma}}} A^{-1}v(x_0,0) \right]
      \nonumber \\  &&  -
      \epsilon \int_0^t e^{\frac{s - t}{\epsilon^{\gamma}}} df(x(s))y(s)
      A^{-1} \eta(s) \, ds
      \nonumber \\    && -
      \int_0^t e^{\frac{s - t}{\epsilon^{\gamma}}} f(x(s))
      A^{-1} \, dW(s)
       \nonumber \\ && +
      \epsilon^{1 -  \gamma} \int_0^t e^{\frac{s - t}{\epsilon^{\gamma}}}
      A^{-1} v(x(s),s) \, ds
                       \nonumber \\ & =: &
 \epsilon J_1(t)  - \epsilon J_2(t) - J_3(t) + \epsilon^{1- \gamma} J_4(t).
\label{e:i3_int_pts}
\end{eqnarray}
Consequently, on account of estimates \eqref{e:ji_est}:
\begin{eqnarray}
\nrm{I_3(t)}{p} \leq C \left( \epsilon^{2p - \sigma} + \epsilon^{(2p-1) \gamma - \sigma} +
\epsilon^{p \gamma - \sigma} \right).
\nonumber
\end{eqnarray}
Thus:
\begin{equation}
\nrm{y(t)}{p}  \leq  C \left( \epsilon^{2p(1-\gamma) - \sigma} + \epsilon^{ -
\gamma - \sigma} +  \epsilon^{-p \gamma - \sigma} \right),
\nonumber
\end{equation}
from which estimate (\ref{eqn:mom_g_s}) follows upon noticing that, for $\gamma \in
(0,2), \,$ $2p(1-\gamma) > -p \gamma$ as well as that $p \geq 1$ \footnote{The presence
of the term $C \, \epsilon^{2 - 2 \gamma}$ in the bound of the second moment of the
particle velocity can become important when studying the problem considered in this
paper numerically. We refer to \cite{paper3_stuart} for details.}. The proof of the
lemma is now complete. \qquad\end{proof} 

\bigskip

From the above lemma we can obtain sharper bound on $I_3(t)$ and $\widehat{J}_2(t)$
defined in \eqref{e:int_hat} for $\gamma \in (0,2)$:
\begin{corollary}
\label{cor:i3_bd_g_s}
Let $x(t)$ be the solution of \eqref{e:motion_1}. Suppose that the conditions of
Lemma \ref{lem:mom_g_s} hold and that $\gamma \in (0,2)$. Then $I_3(t)$ and
$\widehat{J}_2(t)$ satisfy the following estimates
\begin{equation}
\nrm{I_3(t)}{p}  \leq C \, \epsilon^{ \gamma p - \sigma }, \; \; \; \gamma \in (0,
2),
\label{e:i3_bd_g_s}
\end{equation}
and
\begin{equation}
\nrm{\widehat{J}_2(t)}{p}  \leq C \, \epsilon^{- \gamma p - \sigma }, \; \; \;
\gamma \in (0, 2),
\label{e:int_hat_bd_g_s}
\end{equation}
respectively, where $\sigma >0$ is arbitrarily small.
\label{cor:i3}
\end{corollary}
\begin{proof}  Consider the term $J_2(t)$ defined in equation \eqref{e:i3_int_pts}.  Lemmas
\ref{lem:y_int} and \ref{lem:mom_g_s} imply that for $\gamma \in (0,2)$ we have:
\begin{eqnarray}
\nrm{ J_2(t)}{p}    & \leq & C \left(\epsilon^{(p-1)\gamma - \sigma} +
\epsilon^{2(p-1)\gamma - \sigma} \right)
     \nonumber \\ & \leq &
                     C \epsilon^{(p-1)\gamma - \sigma},
\label{e:j2_g_s}
\end{eqnarray}
since $p \geq 1$. We use now \eqref{e:i3_int_pts} and the above estimate, together
with \eqref{e:ji_est} to obtain:
\begin{eqnarray}
\nrm{I_3(t)}{p}    &  \leq &  C \left( \epsilon^{2 p - \sigma} + \epsilon^{2 p - +
(p-1)\gamma -\sigma} + \epsilon^{ p \gamma - \sigma} \right)
                          \nonumber \\ & \leq &
                             C \, \epsilon^{p \gamma - \sigma},
\nonumber
\end{eqnarray}
where we have used the facts that $p \geq 1, \, \gamma <2$. Calculations similar to
the ones used in the proof of Lemma \ref{lem:y_int}, together with estimate
\eqref{eqn:mom_g_s} yield estimate \eqref{e:int_hat_bd_g_s}. \qquad\end{proof} 
%
%
%
%
\section{Drift Corrections To the It\^{o} Integral}
\label{sec:ito_vs_strat}
Estimates \eqref{i1_b}, \eqref{i4_b}, \eqref{i3_g_b} and \eqref{e:i3_bd_g_s}, together with
equation \eqref{e:x_ii} imply that $x(t)$ is of the form:
$$
x(t) = x_0 + I_2(t) + o(1).
$$
Thus, in order to analyze the behavior of $x(t)$ as $\epsilon$ tends to $0$ we need
to identify the contribution of the term $I_2(t)$ to the limiting equation. In this
section we use the bounds on the moments of $y(t)$ that we derived in section
\ref{subsec:prelim_lemmas_y_moments} to investigate precisely the limit of $I_2(t)$
defined by \eqref{term_i_2} as $\epsilon \rightarrow 0$.

We start with the regime $\gamma \in (0,2)$. We have the following:
\begin{lemma}
\label{thm:i2_gamma_small}
Let $x(t)$ be given by \eqref{x_eqn} and let $y(t) = \dot{x}(t)$ be given by
\eqref{y_soln}. Suppose that conditions \eqref{e:b_cond} and \eqref{eqn:conds_g_s}
with $\rho =1$ hold. Then for $\gamma \in (0,2)$ the term $I_2(t)$ given by \eqref{term_i_2}
 has the form:
\begin{equation}
I_2(t) =  \int_0^t f(x(s)) A^{-1} \, d W(s) + H(t)
\nonumber
\end{equation}
where
\begin{equation}
\nrm{H(t)}{p} \leq C \, \epsilon^{(2 - \gamma)p - \sigma},
\label{est:ito_term}
\end{equation}
and where $\sigma > 0$ is arbitrarily small.
\end{lemma}
\begin{proof} $I_2(t)$ is given by equation \eqref{e:i2_thm}:
\begin{eqnarray}
I_2(t)     & = & \int_0^t f(x(s)) A^{-1} \, dW(s) - \epsilon \widehat{J}_1(t)
         + \epsilon \widehat{J}_2(t).
\nonumber
\end{eqnarray}
We have that
\begin{equation}
\nrm{\epsilon \widehat{J}_1(t)}{p} \leq C \, \epsilon^{2p - \sigma}.
\nonumber
\end{equation}
Furthermore, estimate \eqref{e:int_hat_bd_g_s} gives
\begin{equation}
\nrm{\epsilon \widehat{J}_2(t)}{p} \leq C \, \epsilon^{(2- \gamma)p - \sigma}.
\nonumber
\end{equation}
Estimate \eqref{est:ito_term} follows from the above bounds. \qquad\end{proof} 

\bigskip

From our estimates on terms $I_i(t), \, i = 1, \dots 4$ we anticipate that
$x(t)$ converges in mean square, as $\epsilon \rightarrow 0$, to $X(t)$, which
satisfies equation \eqref{limit_ito}. The proof of this convergence is presented in
section \ref{sec:converg_thm}.

Now we proceed with the case $\gamma \in [2,\infty)$. We have the following lemma.
\begin{lemma}
\label{lem:i2_g_b}
Let $x(t)$ be given by (\ref{x_eqn}) and let $y(t) = \dot{x}(t)$ be given by
\eqref{y_soln}. Suppose that conditions \eqref{eqn:conds_efs_g_b}, \eqref{e:b_cond},
\eqref{cond_coeff_h} and \eqref{eqn:conds_g_b} hold. Then for $\gamma
\in [2, \infty)$ the term $I_2(t)$ in (\ref{x_eqn}) has the form:
\begin{eqnarray}
I_2(t) & = & \int_0^t  \nabla \cdot \left( f(x(s)) \Theta f^T(x(s)) \right) \, ds -
\int_0^t   f(x(s)) \Theta \nabla \cdot f^T(x(s))  \, ds
            \nonumber  \\ &&{}
                  + \int_0^t f(x(s))A^{-1} \, d W(s)
            \nonumber \\ &&{}
                  - \epsilon^{\gamma -1} \int_0^t  df(x(s))
          y(s) \eta(s)  \, ds + H(t),
\label{i2_bd_g_b}
\end{eqnarray}
where
$$
\nrm{H(t)}{p} \leq C \, \epsilon^{2p - \sigma}.
$$
where $\sigma > 0$ is arbitrarily small.
\end{lemma}

\begin{proof}  From \eqref{e:i2_thm} we have:
\begin{equation}
I_2(t) = \int_0^t \summ{j} \frac{f_{ij}(x(s))
      \sqrt{\lambda_j}}{\alpha_j} \,
                  d \beta_j(s) - \epsilon \widehat{J}_1(t) + \epsilon \widehat{J}_2(t),
\label{i2_eqn}
\end{equation}
with $\nrm{\epsilon \widehat{J}_1(t)}{p}  \leq  C \, \epsilon^{2 p - \sigma}$
\footnote{This estimate is independent of $\gamma$.}. In order to study the term
$\widehat{J}_2(t)$ we need another two integrations by parts. We apply It\^{o}
formula to the function $$G_{ijk}(x,y, \eta) = f_{ij,k}(x)y_k\alpha_j^{-1}\eta_j$$
from which we obtain, after some algebra:
\begin{eqnarray}
\epsilon \, \frac{f_{ij,k} \eta_j y_k}{\alpha_j} dt
                    & = &
                     -\epsilon^{\gamma+1} d \left(
                     \frac{f_{ij,k} \eta_j y_k}{\alpha_j} \right)
                     +
                     \epsilon^{\gamma +1}
                     \sum_{\ell =1}^d \frac{f_{ij,k\ell} \eta_j y_k
                    y_{\ell}}{\alpha_j} dt
             \nonumber \\   &&
                    + \epsilon \frac{f_{ij,k} \eta_j b_k }{\alpha_j} dt
            + \summ{\rho} \frac{f_{ij,k} f_{k \rho} \eta_j \eta_{\rho}}
                    {\alpha_j} dt
             \nonumber \\   &&
                    - \epsilon^{\gamma} \frac{f_{ij,k}y_k \sqrt{\lambda_j}}{\alpha_j} d \beta_j
                    -  \epsilon^{\gamma-1}f_{ij,k}y_k \eta_j \, dt
\label{eqn:no_sum}
\end{eqnarray}
Now we define the following functions (no summation):
\begin{equation}
F_{ijk \rho} = f_{ij,k}f_{k \rho}
\nonumber
\end{equation}
and
\begin{equation}
G_{ijk \rho} = F_{ijk \rho} \eta_{\rho} \eta_j \alpha_j^{- 1}.
\nonumber
\end{equation}
We apply It\^{o} formula to the function $G_{ijk \rho}$ to obtain:
\begin{eqnarray}
dG_{ijk\rho} & = & \sum_{\ell} F_{ijk \rho, \ell} y_{\ell} \eta_j
                   \eta_{\rho} \alpha_j^{-1} dt
                -
                \frac{1}{\epsilon^2}F_{ijk \rho}\alpha_j^{-1}
                \eta_j \eta_{\rho} (\alpha_{\rho}
                + \alpha_j)dt
             \nonumber \\  && +
                \frac{1}{\epsilon^2}F_{ijk \rho}
                \alpha_j^{-1}  \sqrt{\lambda_j
                \lambda_{\rho}} \delta_{j \rho} dt
                +
                \frac{1}{\epsilon} F_{ijk \rho}
                \alpha_j^{-1}
                ( \eta_{\rho} \sqrt{\lambda_j} d \beta_j +
                \eta_{j} \sqrt{\lambda_{\rho}} d \beta_{\rho}),
\nonumber
\end{eqnarray}
from which, after multiplying through by $\epsilon^2(\alpha_j + \alpha_{\rho})^{-1}$
and taking the sum over $\rho =1,2, \dots$ we get:
\begin{eqnarray}
\summ{\rho} \frac{f_{ij,k} f_{k \rho} \eta_j \eta_{\rho}}{\alpha_j} dt
             &  = &
               \frac{f_{ij,k} f_{kj} \lambda_j}{2 \alpha_j^2} dt
             \nonumber \\ &&
               -
               \epsilon^2 d \left( \summ{\rho} \frac{f_{ij,k}
               f_{k \rho} \eta_{\rho} \eta_{j}}{(\alpha_j + \alpha_{\rho})
           \alpha_j} \right)
            +
               \epsilon^2 \sum_{\ell =1}^d \summ{\rho}
               \frac{(f_{ij,k} f_{k \rho})_{, \ell} y_{\ell} \eta_{\rho}
               \eta_{j}}{\alpha_j(\alpha_{\rho} + \alpha_j)} dt
            \nonumber \\  &&
                 +
                 \epsilon
                 \summ{\rho} \frac{f_{ij,k} f_{k \rho}}{\alpha_j
                 (\alpha_{\rho}+ \alpha_j) }
                ( \eta_{\rho} \sqrt{\lambda_j} d \beta_j +
                \eta_{j} \sqrt{\lambda_{\rho}} d \beta_{\rho}).
\label{eqn:no_sum_2}
\end{eqnarray}
Note that
$$f_{ij,k} f_{kj}=(f_{ij}f_{kj}),k-f_{ij}f_{k,kj}.$$
Thus, from the above calculations, after taking the sum over $j \in \mathbb{Z}^d$
and $k =1, \, \dots, d$ we obtain:
\begin{eqnarray}
\epsilon \widehat{J}_2(t)         & = &
                   \int_0^t  \nabla \cdot \left( f(x(s)) \Theta
                   f^T(x(s)) \right) \, ds - \int_0^t   f(x(s)) \Theta \nabla \cdot
                   f^T(x(s))  \, ds
            \nonumber  \\ &&{}
                  + \int_0^t f(x(s))A^{-1} \, d W(s)
          - \epsilon^{\gamma -1} \int_0^t  df(x(s)) y(s) \eta(s)  \, ds
                    + H(t),
\nonumber
\end{eqnarray}
where $H(t) = \sum_{\ell = 1}^8 H_{\ell}(t)$ with
\begin{subequations}
\begin{equation}
H_1^i(t) = - \left. \epsilon^2 \summ{j, \rho} \sum_{k =
        1}^d \frac{f_{ij,k}(x(s)) f_{k \rho}(x(s)) \eta_{\rho}(s)
        \eta_{j}(s)}{\alpha_j (\alpha_{\rho} + \alpha_j)} \right|_0^t,
%
%
\end{equation}
\begin{equation}
H_2^i(t) =  \epsilon^2 \int_0^t \summ{j, \rho}
          \sum_{k, \ell = 1}^d
          \frac{(f_{ij,k \ell}(x(s)) f_{k \rho}(x(s)))_{,\ell} y_{\ell}(s)
      \eta_{\rho}(s)
          \eta_{j}(s)}{\alpha_j(\alpha_j + \alpha_{\rho})} \, ds,
%
%
\end{equation}
\begin{equation}
H_3^i(t) =  \epsilon \int_0^t \summ{j, \rho}
          \sum_{k = 1}^d \frac{f_{ij,k}(x(s)) f_{k \rho}(x(s))
          \eta_{\rho}(s) \sqrt{\lambda_{j}}}{\alpha_j(\alpha_j +
          \alpha_{\rho})} \, d \beta_j(s),
%
%
\end{equation}
\begin{equation}
H_4^i(t) =  \epsilon \int_0^t \summ{j, \rho}
          \sum_{k = 1}^d \frac{f_{ij,k}(x(s)) f_{k \rho}(x(s))
          \eta_{j}(s) \sqrt{\lambda_{\rho}}}{\alpha_j(\alpha_j +
          \alpha_{\rho})} \, d \beta_{\rho}(s),
%
%
\end{equation}
\begin{equation}
H_5^i(t) = - \left. \epsilon^{\gamma +1} \summ{j} \sum_{k =
        1}^d \frac{f_{ij,k}(x(s)) y_k(s) \eta_{j}(s)}{\alpha_j} \right|_0^t,
%
%
\end{equation}
\begin{equation}
H_6^i(t) =  \epsilon^{\gamma +1} \int_0^t \summ{j}
          \sum_{k, \ell = 1}^d
          \frac{f_{ij,k \ell}(x(s)) y_k(s) y_{\ell}(s) \eta_{j}(s)}
      {\alpha_j} \, ds,
%
%
\end{equation}
\begin{equation}
H_7^i(t) =  \epsilon \int_0^t \summ{j}
          \sum_{k = 1}^d \frac{f_{ij,k}(x(s)) \eta_{j}(s) b_k(x(s))}
      {\alpha_j} \, ds,
%
%
\end{equation}
\begin{equation}
H_8^i(t) =  \epsilon^{\gamma} \int_0^t \summ{j}
          \sum_{k = 1}^d \frac{f_{ij,k}(x(s))  y_k(s) \sqrt{\lambda_j}}
      {\alpha_j} \, d \beta_j(s).
%
%
\end{equation}
\label{e:h}
\end{subequations}
Now we have to bound the terms $H_{\ell}(t), \, \ell = 1, \dots 8$. The necessary
estimates are proved in Lemma \ref{lem:h} which is presented in the Appendix. The final
result is that
$$
\nrm{ H(t)}{p} \leq C \epsilon^{2p - \sigma}.
$$
This completes the proof of the lemma. \qquad\end{proof} 

\bigskip

Lemma \ref{lem:i2_g_b} together with the estimates on $I_3(t), \, I_4(t)$ will enable
us to show that for $\gamma \in (2, \infty)$ the particle position $x(t)$ converges
in $L^{2p}(\Omega, C([0,T]; \mathbb{R}))$ to the solution of the It\^{o} SDE
(\ref{limit_stratonovich_1}). The precise convergence theorem will be proved in the
next section.

The above argument fails when $\gamma = 2$ since in this case we cannot control the
last integral on the right hand side of \eqref{i2_bd_g_b}
uniformly in $\epsilon$. For the convergence theorem in this case we need the
following corollary of the previous lemma:
\begin{corollary}
\label{cor:i2_g_2}
Let the conditions of Lemma \ref{lem:i2_g_b} be satisfied and let $\gamma = 2$. Then
$I_2(t)$ has the following form:
\begin{eqnarray}
I_2(t) & = & \int_0^t  \nabla \cdot \left( f(x(s)) \widehat{\Theta} f^T(x(s))
\right) \, ds - \int_0^t   f(x(s)) \widehat{\Theta} \nabla \cdot f^T(x(s))  \, ds
            \nonumber  \\ &&{}
                 + \int_0^t f(x(s))A^{-1} \, d W(s)  + \widehat{H}(t),
\label{eqn:i2_g_2}
\end{eqnarray}
with $\nrm{\widehat{H}(t)}{p} \leq \epsilon^{2 p - \sigma}$,
where $\sigma > 0$ is arbitrarily small.
\end{corollary}
\begin{proof} We set $\gamma = 2$ in \eqref{eqn:no_sum}, solve for $f_{ij,k} y_k \eta_j
\alpha_j^{-1} dt$ and combine the result with \eqref{eqn:no_sum_2}, sum over $j \in
\mathbb{Z}^d$ and $k = 1, \dots, d$ and integrate over $[0,t]$ to obtain
\begin{eqnarray}
I_2(t) &=& \int_0^t  \nabla \cdot \left( f(x(s)) \widehat{\Theta} f^T(x(s)) \right)
\, ds - \int_0^t   f(x(s)) \widehat{\Theta} \nabla \cdot f^T(x(s))  \, ds
            \nonumber  \\ &&{}
                 + \int_0^t f(x(s))A^{-1} \, d W(s)  + \widehat{H}(t),
\nonumber
\end{eqnarray}
with $\widehat{H}(t) = \sum_{k = 1}^8 \widehat{H}_k(t)$. The terms
$\widehat{H}_k(t), \, k =1, \dots, 8$ are similar to the terms defined in
\eqref{e:h}, with the difference that the diagonal operator $(I + A)^{-1}$ is
applied to every one of them. Similar techniques to the ones used in the proof of
the previous ones still apply. We obtain estimate $\nrm{\widehat{H}(t)}{p} \leq C \,
\epsilon^{2p - \sigma}$, provided that conditions \eqref{eqn:conds_g_b} hold.  \qquad\end{proof} 
\bigskip

The above corollary will enable us to show that, for $\gamma =2$, $x(t)$ converges to
$X(t)$ which satisfies SDE (\ref{limit_stratonovich_3}). This leads to the
surprising conclusion that in this case the correction to the drift is not the usual
Stratonovich correction. The precise convergence theorem will be proved in the next
section.
%
%
\section{Proofs of The Convergence Theorems}
\label{sec:converg_thm}
In this section we prove the convergence Theorems \ref{thm:gamma_small},
\ref{thm:gamma_big} and \ref{thm:gamma_two}. In the following proofs we will use the
fact that $f: \ell_2 \rightarrow \R^d$ is a Lipschitz continuous map, provided that
condition \eqref{eqn:conds_g_s_2} with $\rho = \frac{1}{2}$ holds. In particular, a
calculation similar to the one presented in the proof of Lemma \ref{lemma:stoch_conv},
equation \eqref{e:trace}  yields:
\begin{equation}
\|\left( f(X(s)) - f(x(s)) \right)A^{-1} \|_{L^2_0} \leq C \, \|X(s) - x(s) \|.
\label{eqn:lip_conts}
\end{equation}

{\it Proof of Theorem\/ {\rm\ref{thm:gamma_small}}}. We combine \eqref{x_eqn} together with
estimates \eqref{i1_b}, \eqref{i4_b}, Corollary \ref{cor:i3_bd_g_s}
and Lemma \ref{thm:i2_gamma_small} to write $x(t)$ in the form:
\begin{equation}
x(t) = x_0 +  \int_0^t f(x(s)) A^{-1} \, dW(s) + \int_0^t b(x(s)) \, ds + R_1,
\label{e:x_g_s}
\end{equation}
with  $\nrm{R_1}{p} \leq C \left( \epsilon^{(2 - \gamma)p - \sigma} +
\epsilon^{\gamma p -\sigma} \right)$.
Now we take the difference between $X(t)$ given by \eqref{limit_ito} and $x(t)$
given by \eqref{e:x_g_s}, raise it to the $2p$--th power, take the expectation value
of the supremum, use the Burkholder--Davis--Gundy inequality and the Lipschitz continuity of
$f(x), \, b(x)$, together with the estimate on $R_1$ to obtain:
\begin{eqnarray}
\nrm{X(t) - x(t)}{p}   & \leq &  C \int_0^T  \E \| (f(X(s)) - f(x(s)))A^{-1}
                        \|^{2p}_{L^0_2} \, ds
                       \nonumber \\ && + C \int_0^T \mathbb{E} \left( \sup_{0 \leq t
                        \leq s}  \|X(t) - x(t) \|^{2p} \right) \, d s
                        +  C \, \nrm{R_1(t)}{p}
                 \nonumber \\ & \leq &
                            C \left( \epsilon^{(2 - \gamma)p - \sigma}
                            + \epsilon^{\gamma p -\sigma} \right) +  C \int_0^T
                             \mathbb{E} \left( \sup_{0 \leq t \leq s}
                             \|X(t) - x(t) \|^{2p} \right) \, d s.
\nonumber
\end{eqnarray}
We apply now Gronwall's lemma to the above equation for the function $$\xi(T) =
\nrm{X(t) -x(t)}{p}$$ to conclude the proof of Theorem \ref{thm:gamma_small}. \qquad\endproof 

\bigskip

Now we proceed with the convergence theorems for $\gamma \in [2,\infty)$. Let us
consider the case $\gamma > 2$.

{\it Proof of Theorem\/ {\rm\ref{thm:gamma_big}}} . Consider the integral
$$
J(t) = \epsilon^{\gamma -1} \int_0^t df(x(s)) y(s) \eta(s) \, ds.
$$
We use Lemma \ref{lem:y_int} with $\rho = 0$ to deduce
$$
\nrm{J(t)}{p} \leq C \epsilon^{2p(\gamma - 2) - \sigma}.
$$

Now we combine \eqref{x_eqn} with estimates \eqref{i1_b},
\eqref{i4_b}, \eqref{i3_g_b} and Lemma \ref{lem:i2_g_b},
together with the above estimate, to write $x(t)$ in the form:
\begin{eqnarray}
x(t) & = & x_0 + \int_0^t B(x(s)) \, ds + \int_0^t f(x(s)) A^{-1} \, d W(s) +
R_2(t),
\label{eqn:x_g_b}
\end{eqnarray}
with $\normm{R_2(t)} \leq C \, \left( \epsilon^{2p (\gamma - 2) - \sigma} +
\epsilon^{2p - \sigma} \right)$ and
\begin{equation}
B(x(s)) =  b(x(s)) +  \nabla \cdot \left( f(x(s)) \Theta f^T(x(s)) \right) - f(x(s))
\Theta \nabla \cdot f^T(x(s)).
\nonumber
\end{equation}
Now, assumptions \eqref{e:b_cond} and \eqref{eqn:conds_efs_g_b}
imply that the drift term $B(x)$ in equation \eqref{eqn:x_g_b} is Lipschitz continuous:
\begin{equation}
\left\| B(X) - B(x) \right\| \leq C \, \|X - x \|,
\label{eqn:lip_drift}
\end{equation}
under condition \eqref{eqn:conds_g_b_2}. The Lipschitz continuity of $B(x)$,
together \eqref{eqn:lip_conts} and the Burkholder--Davis--Gundy inequality give
\begin{equation}
\nrm{X(t) - x(t)}{p} \leq C_1 \left( \epsilon^{2p (\gamma - 2) - \sigma} +
\epsilon^{2p - \sigma} \right) + C_2 \int_0^T \mathbb{E}\left( \sup_{0 \leq t \leq
s} \| X(t) - x(t) \|^{2p} \right) \, ds,
\nonumber
\end{equation}
from which Theorem \ref{thm:gamma_big} follows, upon applying Gronwall's lemma. \qquad\endproof 

\bigskip

Now we are ready to present the convergence proof and theorem for the case $\gamma
=2$. Since the proof is essentially the same as the one of Theorem
\ref{thm:gamma_big}, we will be brief.

{\it Proof of Theorem\/ {\rm\ref{thm:gamma_two}}}. We combine \eqref{x_eqn} with estimates
\eqref{i1_b}, \eqref{i4_b}, \eqref{i3_g_b} and Corollary
\ref{cor:i2_g_2}, to write $x(t)$ in the form--for $\gamma = 2$:
\begin{eqnarray}
x(t) & = & x_0 + \int_0^t \widehat{B}(x(s)) \, ds + \int_0^t f(x(s)) A^{-1} \, d
W(s) + R_3(t),
\label{e:x_g_2}
\end{eqnarray}
with $\nrm{ R_3(t)}{p} \leq  C \, \epsilon^{2p-\sigma} $ and
\begin{equation}
\widehat{B}(x(s)) =  b(x(s)) +  \nabla \cdot \left( f(x(s)) \widehat{\Theta}
f^T(x(s)) \right) - f(x(s)) \widehat{\Theta} \nabla f^T(x(s)).
\nonumber
\end{equation}
Assumptions assumptions \eqref{e:b_cond}, \eqref{eqn:conds_efs_g_b} and
\eqref{eqn:conds_g_b_2} and ensure that $\widehat{B}(x)$ is Lipschitz continuous. As in the
proof of the previous theorem, we take the difference between $X(t)$ given by
\eqref{limit_stratonovich_3} and $x(t)$ given by \eqref{e:x_g_2}), raise it to the
$2p$-th power, take the expectation value of the supremum, use the
Burkholder--Davis--Gundy inequality and the Lipschitz continuity of the terms in
\eqref{e:x_g_2} to obtain.
\begin{equation}
\nrm{X(t) - x(t)}{p} \leq C_1 \, \epsilon^{2p - \sigma} + C_2 \int_0^T \nrm{ X(t) -
x(t)}{p} \, ds,
\nonumber
\end{equation}
Now we apply Gronwall's lemma to obtain estimate \eqref{etimate_thm_gamma_two}. \qquad\endproof 
%
%
\section{Applications}
\label{sec:applications}
\subsection{Inertial Particles in a Random Field}
A model for the motion of inertial particles in turbulent flows was introduced in
\cite{inertial_1, inertial_2}. It consists of Stokes' law for the particle motion
with the background divergence--free fluid velocity field being an infinite
dimensional Ornstein--Uhlenbeck process. We assume that the motion takes place on
the two--dimensional unit torus $\T^2$:
\begin{subequations}
\begin{equation}
\tau \ddot{x} = v(x,t) - \dot{x},
\end{equation}
\begin{equation}
v = \nabla^{\bot} \psi,
\end{equation}
\begin{equation}
d \psi = \nu \Delta \psi \, dt + \sqrt{\nu} dW,
\label{e:ou_appl}
\end{equation}
\label{e:in_part_appl}
\end{subequations}
where $\nabla^{\bot} := (\frac{\partial}{\partial x_2}, -\frac{\partial}{\partial
x_1})^T$ stands for the skew--gradient and $\psi$ denotes the stream function.
Furthermore, $W(x,t)$ denotes a $Q$--Wiener process on $$H := \left\{ f \in
L^2_{per}(\T^2); \int_{\T^2} f dx =0 \right\}.$$ Various asymptotic limits for
\eqref{e:in_part_appl} were considered in \cite{paper1_stuart}. Let us consider now
the scaling limit considered in this paper.

We assume that $\tau = \tau_0 \eps^{\gamma -1}$, that the inverse noise correlation
time $\nu$ is of $\mathcal{O}(\eps^{-1})$ and rescale time by $t \rightarrow
t/\eps$. Moreover, we expand the solutions of \eqref{e:ou_appl} in terms of the
eigenfunctions of the Laplacian on $\T^2$, $e^{i k \cdot x}$. We also set $K = 2 \pi
\mathbb{Z}^2 \setminus \{(0,0) \}$ and denote $\widehat{\mathbb{C}}^K := \{\eta \in
\mathbb{C}^K; \eta_k = \bar{\eta}_{-k} \}$, equipped with the standard $\ell_2$
inner product. Setting $\tau_0 =1$ for notational simplicity, the rescaled equations
\eqref{e:in_part_appl} can be written in the form:
\begin{subequations}
\begin{equation}
\epsilon^{\gamma} \, \ddot{x} = \frac{f(x) \eta(t)}{\epsilon}  - \dot{x}(t),
\label{e:motion_resc}
\end{equation}
\begin{equation}
f(x) \xi = \sum_{k \in K} i k^{\bot} e^{i k \cdot x} \xi_k,
\end{equation}
\begin{equation}
d \eta_k = - \frac{1}{\epsilon^2} |k|^2 \eta_k dt + \frac{1}{\epsilon}
\sqrt{\lambda_k} d \beta_k, \; \; k \in K.
\label{e:ou_in_part}
\end{equation}
\end{subequations}
with $k^{\bot} = [k_2  \; -k_1]^T$, $\eta = \{ \eta_k \}_{k =1}^{\infty} \in
\widehat{\mathbb{C}}^K$. Moreover, $\{\beta_k(t) \}_{k =1}^{\infty}$ are mutually
independent one dimensional standard Brownian motions satisfying the reality conditions
$\beta_k = \bar{\beta}_{-k}$.

It was shown in \cite{inertial_1} that
\begin{equation}
 f \Theta f^T = \sigma I \quad \mbox{with} \quad \sigma = \sum_{k \in K}
 \frac{\lambda_k}{2 |k|^2},
\label{e:kubo}
\end{equation}
where $\Theta$ is defined in equation \eqref{e:chi} and $I$ stands for the identity
matrix. A similar calculation reveals:
$$
 f \widehat{\Theta} f^T = \sigma I \quad \mbox{with} \quad \widehat{\sigma} = \sum_{k \in K}
 \frac{\lambda_k}{2 |k|^2(1 + |k|^2)}.
$$
Furthermore, the incompressibility of the velocity field implies that
$$
\nabla \cdot f^T =0.
$$
The above calculations imply that, for the inertial particles problem whose motion
is modelled by \eqref{e:in_part_appl}, the It\^{o} and Stratonovich interpretations
of the stochastic integral coincide and that the limiting equation of motion is
$$
X(t) = x_0 + \int_0^t f(X(s)) A^{-1} \, dW(s),
$$
for all values of $\gamma >0$. In fact, Theorems \ref{thm:gamma_small},
\ref{thm:gamma_big} and \ref{thm:gamma_two}, together with the properties of the
eigenfunctions of the Laplacian on $\T^2$ yield
\begin{proposition}
Let $x(t)$ be the solution of \eqref{e:motion_resc} and $\gamma \in (0, \infty)$.
Assume that
$$
\sum_{k \in K} \sqrt{\lambda_k} < \infty, \quad \mbox{for} \quad \gamma \in (0,2)
$$
and that
$$
\sum_{k \in K} \sqrt{\lambda_k} |k| < \infty, \quad \mbox{for} \quad \gamma \in [2,
\infty).
$$
Assume further that conditions \eqref{e:ic_g_s} \eqref{e:ic_g_b} hold and that the initial
conditions for \eqref{e:ou_in_part} are stationary. Then $x(t)$
converges, as $\eps \rightarrow 0$, to $X(t)$ which satisfies
$$
X(t) = x_0 + \int_0^t f(X(s)) A^{-1} \, dW(s),
$$
with
$$
\nrm{X(t) - x(t)}{p} \leq C \left(\eps^{\gamma p} + \eps^{(2-\gamma)p - \sigma}
\right) \quad \mbox{for} \quad \gamma \in (0,2),
$$
$$
\nrm{X(t) - x(t)}{p} \leq C \eps^{2 - \sigma} \quad \mbox{for} \quad \gamma = 2
$$
and
$$
\nrm{X(t) - x(t)}{p} \leq C \left(\eps^{ 2p - \sigma} + \eps^{2p(\gamma -2) -
\sigma} \right) \quad \mbox{for} \quad \gamma \in (2, \infty),
$$
where $\sigma >0$ is arbitrarily small. The constant $C$ depends on the moments of the
initial conditions, the spectrum of the Wiener process, the operator $A$, the
exponent $p$, the maximum time $T$ and $\sigma$.
\end{proposition}
Physically we are looking at inertial particles in rapidly decorrelating velocity
fields, over long times. The parameter $\gamma$ effects the non--dimensional mass of
the particle, which is of $\mathcal{O}(\eps^{\gamma -1})$. Provided that $\gamma
>0$, which includes a range of massive as well as light particles, the limiting
particle motion is equivalent in law to a Brownian motion, see eq. \eqref{e:kubo}.
For $\gamma= 0$, however, the limiting motion is that of the integrated OU process:
the particle velocity is of OU type \cite{paper1_stuart}.
\subsection{Diffusion in Solids}
Consider now the motion of a particle in one dimension under the influence of a periodic
potential $V(x)$, subject to dissipation:
\begin{equation}
\tau \ddot{x} = - V'(x) - \dot{x}.
\label{e:motion}
\end{equation}
We assume that the derivative of the potential can be written in the following Fourier sine
series:
$$
V'(x) = -\sum_{j=1}^{\infty} \sin(jx) \mu_j.
$$
We assume further that the {\it control parameters} $\mu_j$ are noisy and of the form
$$
\mu_j = \mu_j^0 + \frac{1}{\eps}\eta_j(t/\eps^2),
$$
where $\{ \mu^0_j\}_{j =1}^{\infty}$  are constants and $\{\eta_j(t) \}_{j=1}^{\infty}$ are
one--dimensional OU processes driven by mutually independent noises:
$$
d \eta_j = - j^2 \eta_j dt + \sqrt{\lambda_j} d \beta_j, \; \; j = 1 \dots \infty.
$$
Substituting the above into \eqref{e:motion} and assuming that the particle relaxation time
$\tau$  is of $\mathcal{O}(\eps^{\gamma})$ we obtain
\begin{subequations}
\begin{equation}
\epsilon^{\gamma} \ddot{x} = - V_0'(x) - \dot{x} + \frac{1}{\eps} \sum_{j
=1}^{\infty} \sin(jx) \eta_j(t)
\end{equation}
\begin{equation}
d \eta_j = - \frac{1}{\epsilon^2} j^2 \eta_j dt + \frac{1}{\epsilon}
\sqrt{\lambda_j} d \beta_j, \; \; j = 1 \dots \infty,
\nonumber
\end{equation}
\label{e:mol_appl}
\end{subequations}
where $V_0'(x) = -\sum_{j=1}^{\infty} \sin(jx) \mu^0_j.$ We use now Theorems
\ref{thm:gamma_small}, \ref{thm:gamma_big} and  \ref{thm:gamma_two}, to deduce the
following result.
\begin{proposition}
Let $x(t)$ be the solution of \eqref{e:mol_appl} and $\gamma \in (0, \infty)$.
Assume that
$$
\sum_{j =1}^{\infty} \sqrt{\lambda_j} j^{-1} < \infty, \quad \mbox{for} \quad \gamma
\in (0,2)
$$
and that
$$
\sum_{j =1}^{\infty} \sqrt{\lambda_j}  < \infty, \quad \mbox{for} \quad \gamma \in
[2, \infty).
$$
Assume further that conditions \eqref{e:ic_g_s} \eqref{e:ic_g_b} hold, that $V_0(x) \in
C_b^2(\R)$ and that the initial conditions for \eqref{e:ou_in_part} are stationary. Then $x(t)$
converges, as $\eps \rightarrow 0$, to $X(t)$ which satisfies
\begin{eqnarray}
X(t) = \left\{ \begin{array}
            {r@{\quad:\quad} l}
            x_0 - V_0'(x) + \int_0^t \sum_{j=1}^{\infty} \frac{\sqrt{\lambda_j}}{j^2}
        \sin(j X(s)) \, d \beta_j(s)
            & \gamma < 2 \; \; \;
             \medskip

               \\
             x_0 - V_0'(x) + \frac{1}{4} \int_0^t  \sum_{j=1}^{\infty} \frac{\lambda_j}
             {j^3(1 +j^2)} \sin(2j X(s)) \, ds +  \int_0^t \sum_{j=1}^{\infty}
             \frac{\sqrt{\lambda_j}}{\alpha_j} \sin( j X(s)) \, d \beta_j(s)
             & \gamma = 2
         \medskip
             \\
%
%
            x_0 - V_0'(x) + \frac{1}{4} \int_0^t  \sum_{j=1}^{\infty} \frac{\lambda_j}{j^3}
        \sin(2 j X(s)) \, ds + \int_0^t \sum_{j=1}^{\infty} \frac{\sqrt{\lambda_j}}{j^2}
        \sin(j X(s)) \, d \beta_j(s)
             & \gamma > 2 \; \; \;
            \end{array}   \right.  .
\nonumber
\end{eqnarray}
with
\begin{equation}
\nrm{X(t) - x(t)}{p} \leq C \left(\eps^{\gamma p} + \eps^{(2-\gamma)p - \sigma}
\right) \quad \mbox{for} \quad \gamma \in (0,2),
\nonumber
\end{equation}
\begin{equation}
\nrm{X(t) - x(t)}{p} \leq C \eps^{2p - \sigma} \quad \mbox{for} \quad \gamma =2
\nonumber
\end{equation}
and

$$
\nrm{X(t) - x(t)}{p} \leq C \left(\eps^{ 2p - \sigma} + \eps^{2p(\gamma -2) -
\sigma} \right) \quad \mbox{for} \quad \gamma \in [2, \infty),
$$
where $\sigma >0$ is arbitrarily small. The constant $C$ depends on the moments of the
initial conditions, the spectrum of the Wiener process, the operator $A$, the
exponent $p$, the maximum time $T$ and $\sigma$.
\end{proposition}
We remark that for $\gamma \in (0,2)$ the particle motion is in the mean potential
$$
V_0(x) = \E (V(x,t)).
$$
On the other hand, for $\gamma \geq 2$, the limiting motion particle motion is in
modified potential which depends discontinuously on $\gamma$ as $\gamma \rightarrow
2^+$.
\section{Conclusions}
\label{sec:conclusions}
The It\^{o} versus Stratonovich problem is studied in this paper for a class of infinite
dimensional mean zero Gaussian random fields. It is shown that the correct interpretation of
the stochastic integral in the limiting equation depends on the rate with which the particle
relaxation time $\tau_p$ tends to $0$, relative to that of the noise correlation time $\tau_n$.
In particular, it was shown that in the case where $\tau_p$ and $\tau_n$ tend to zero at the
same rate, then the limiting stochastic integral in neither of It\^{o} nor of Stratonovich type.

The proof of our convergence theorems is based entirely on the pathwise techniques
developed in \cite{dowell} and used previously in \cite{paper1_stuart}, rather than
the weak convergence methods of e.g. \cite{ethier_86}. Our techniques enable us to
obtain strong, i.e. pathwise, convergence results as well as sharp upper bounds on
the convergence rates. A drawback of the method employed in this paper is that it is
applicable only for noise processes which can be expressed as solutions of
stochastic differential equations, like the one used in this paper. In order to
apply the results reported in this paper to more general classes of colored
approximations to white noise, weak convergence techniques will be more appropriate.
%
%
%
{\appendix
\section{Estimates on terms $H_i(t), \, i = 1, \dots 8$.}
\label{app:hi_est}
In this appendix we prove the following lemma.
\begin{lemma}
\label{lem:h}
Consider the terms $H_i(t), \, i = 1, \dots 8$ defined in \eqref{e:h} and set $H(t) =
\sum_{i = 1}^8 H_i(t)$. Assume that conditions \eqref{eqn:conds_efs_g_b},
\eqref{eqn:conds_g_b_4} \eqref{e:b_cond},
\eqref{cond_coeff_h}, \eqref{eqn:conds_g_s_1} with $\rho = \frac{1}{2}$ and
\eqref{eqn:conds_g_s_2} with $\rho = 0$  hold. Then the following estimate holds:
\begin{equation}
\nrm{ H(t)}{p} \leq C \epsilon^{2p}.
\label{e:h_app}
\end{equation}
\end{lemma}
\begin{proof} We start with $H_1(t)$. First we compute:
\begin{eqnarray}
\left| \summ{j, \rho} \sum_{k = 1}^d \frac{f_{ij,k}(x(s)) f_{k \rho}(x(s))
\eta_{\rho}(s)  \eta_{j}(s)}{\alpha_j (\alpha_{\rho} + \alpha_j)} \right|
           & \leq & C \summ{\rho} \frac{\alpha_\rho^{\alpha + r}|\eta_{\rho}(s)|}{\alpha_{\rho}}
                   \summ{j} \frac{\alpha_j^{\beta + r}|\eta_j(s)|}{\alpha_j}
        \nonumber \\ & \leq &
         C  \left( \summ{\rho}\alpha_\rho^{\alpha + r-1} |\eta_{\rho}(s)|
           \right)^2 + C \left( \summ{j} \alpha_j^{\beta + r-1}|\eta_j(s)|
           \right)^2  \nonumber \\ &=:& J_1(t) + J_2(t).
\nonumber
\end{eqnarray}
Now, calculations similar to the ones employed in the proof of Lemma
\ref{lem:est_vel} enable us to obtain
$$
\nrm{J_1(t)}{p} \leq C \epsilon^{-\sigma}, \qquad  \nrm{J_2(t)}{p} \leq C
\epsilon^{-\sigma},
$$
for $\sigma > 0$ is arbitrarily small, provided that conditions
\eqref{eqn:conds_g_s_2} with $\rho = 1$ and \eqref{eqn:conds_g_s_1} with $\rho = 1$
hold. We use the above estimate and the definition of $H_1(t)$ to conclude:
$$
\nrm{H_1(t)}{p} \leq C \epsilon^{4p-\sigma},
$$
for $\sigma > 0$ is arbitrarily small.

We proceed now with $H_2(t)$. We define:
$$
J^i(t):= \summ{j, \rho}  \sum_{k, \ell = 1}^d
          \frac{(f_{ij,k \ell}(x(s)) f_{k \rho}(x(s)))_{,\ell} y_{\ell}(s)
      \eta_{\rho}(s)
          \eta_{j}(s)}{\alpha_j(\alpha_j + \alpha_{\rho})}.
$$
Now we compute
\begin{eqnarray}
J^i(t)    & = &
 \summ{j, \rho}  \sum_{k, \ell = 1}^d
          \frac{f_{ij,k \ell \ell}(x(s)) f_{k \rho}(x(s)) y_{\ell}(s)
      \eta_{\rho}(s)
          \eta_{j}(s)}{\alpha_j(\alpha_j + \alpha_{\rho})}
        +
          \summ{j, \rho}  \sum_{k, \ell = 1}^d
          \frac{f_{ij,k \ell}(x(s)) f_{k \rho , \ell}(x(s)) y_{\ell}(s)
      \eta_{\rho}(s)
          \eta_{j}(s)}{\alpha_j(\alpha_j + \alpha_{\rho})}
          \nonumber \\ & \leq &
       C \summ{j} \frac{\alpha_j^{\delta + r} \|y(s) \| | \eta_{j}(s) |}{\alpha_j}
        \summ{\rho} \frac{\alpha_{\rho}^{\alpha + r} | \eta_{\rho}(s)
        |}{\alpha_{\rho}} +
        C \summ{j} \frac{\alpha_j^{\gamma + r} \|y(s) \| | \eta_{j}(s) |}{\alpha_j}
        \summ{\rho} \frac{\alpha_\rho^{\beta + r}  | \eta_{\rho}(s) |}{\alpha_{\rho}}
          \nonumber \\ & \leq &
       C \epsilon^{\zeta} \|y(s) \|^2 \left[ \left( \summ{j} \alpha_j^{\delta + r-1}
       | \eta_{j}(s) | \right)^2 + \left( \summ{j} \alpha_j^{\gamma + r - 1}
       | \eta_{j}(s) | \right)^2 \right]
        \nonumber \\ && +
        C \epsilon^{-\zeta} \left[ \left( \summ{\rho} \alpha_{\rho}^{\alpha + r-1}
        | \eta_{\rho}(s) | \right)^2 +
         \left( \summ{\rho} \alpha_{\rho}^{\beta + r - 1}
        | \eta_{\rho}(s) | \right)^2 \right],
\nonumber
\end{eqnarray}
for $\zeta \in \R$. We use now calculations similar to those used in order to prove
Lemma \ref{lem:est_vel} and estimate \eqref{e:bootstrap}, together with Lemma
\ref{lem:mom_g_b} to deduce
$$
\nrm{H_2(t)}{p} \leq C (\epsilon^{2p \zeta - \sigma} + \epsilon^{4p -2p \zeta -
\sigma}),
$$
provided that conditions  \eqref{eqn:conds_g_s} with $\rho = 1$ hold. We choose now $\zeta =1$
to obtain
$$
\nrm{H_2}{p} \leq C \, \epsilon^{2p - \sigma}.
$$
Consider now the term $H_4(t)$. We introduce the cylindrical Wiener process
$$
\widehat{W}(t) = \summ{\rho} \widehat{e}_k \beta_k(t).
$$
Now we can write $H_4(t)$ in the form
$$
H_4(t) = \epsilon \int_0^t \widehat{F} \, d \widehat{W},
$$
where the map $\widehat{F}: \ell_2 \rightarrow \R^d$ is defined as
$$
\left\{ \widehat{F} \gamma \right\}_i =  \summ{j, \rho}
          \sum_{k = 1}^d \frac{f_{ij,k}(x(s)) f_{k \rho}(x(s))
          \eta_{j}(s) \sqrt{\lambda_{\rho}}}{\alpha_j(\alpha_j +
          \alpha_{\rho})} \gamma_{\rho}, \quad i = 1, \dots d, \quad \forall \gamma \in \ell_2.
$$
We need to estimate the Hilbert--Schmidt norm of $\widehat{F}$. We have:
\begin{eqnarray}
\|\widehat{F} \|^2_{L_2(\ell_2 , \R^d)} & = & \summ{ \rho} \sum_{i = 1}^d
          \left| \sum_{k = 1}^d \summ{j} \frac{f_{ij,k}(x(s)) f_{k \rho}(x(s))
          \eta_{j}(s) \sqrt{\lambda_{\rho}}}{\alpha_j(\alpha_j +
          \alpha_{\rho})} \right|^2
         \nonumber \\ & \leq &
           C \summ{\rho} \left( \summ{j} \frac{\alpha_j^{\beta + r} \alpha_{\rho}^{\alpha + r}
       |\eta_j| \sqrt{\lambda_{\rho}}}{\alpha_j (\alpha_j + \alpha_{\rho})} \right)^2
         \nonumber \\ & \leq &
          C \summ{\rho} \left( \alpha_{\rho}^{\alpha + r - 1}  \sqrt{\lambda_{\rho}}
          \right)^2  \left(\summ{j}  \alpha_j^{\beta + r - 1} |\eta_j| \right)^2
          \nonumber \\ & \leq & C \left(\summ{j}  \alpha_j^{\beta + r - 1} |\eta_j|
            \right)^2,
\nonumber
\end{eqnarray}
provided that condition \eqref{eqn:conds_g_s_1} with $\rho = \frac{1}{2}$ holds. We
use now the Burkholder--Davis--Gundy and H\"{o}lder  inequalities, together with a
calculation similar to the one used in the proof of Lemma \ref{lem:est_vel} to
deduce:
\begin{eqnarray}
\nrm{H_4(t)}{p} & \leq & C \epsilon^{2p} \int_0^T \E \|\widehat{F} \|_{L_2(\ell_2,
\R^d)}^{2p}  \, ds \leq C \epsilon^{2p},
\end{eqnarray}
provided that condition \eqref{eqn:conds_g_s_2} with $\rho = 1$ holds. Exactly the
same analysis provides us with the estimate
$$
\nrm{H_3(t)}{p}  \leq C \epsilon^{2p},
$$
under conditions \eqref{eqn:conds_g_s_2} with $\rho = \frac{1}{2}$ and
\eqref{eqn:conds_g_s_1} with $\rho = 1$.

Now we consider term $H_5(t)$. We have
\begin{equation}
H_5(t) = - \epsilon^{\gamma +1} df(x(s)) y(s) A^{-1} \eta(s) \Big|_0^t.
\nonumber
\end{equation}
Now, the calculations used in the proof of Lemma \ref{lem:y_int}, together with
\eqref{e:bootstrap} yield
\begin{eqnarray}
\nrm{H_5(t)}{p} & \leq & C \, \epsilon^{2p (\gamma +1)} \nrm{df(x(s)) y(s) A^{-1} \eta(s)}{p}
                  \nonumber \\ & \leq &
                   C \, \epsilon^{2 \gamma p - \sigma},
\end{eqnarray}
for $\sigma >0$ is arbitrarily small, provided that \eqref{eqn:conds_g_s_2} holds with $\rho = 1$.

Now we proceed with $H_6(t)$. We use a simple variant of \eqref{e:bootstrap} and
assume \eqref{eqn:conds_g_b_4} to compute:
\begin{eqnarray}
\nrm{H_6(t)}{p} & \leq & C \epsilon^{2p(\gamma + 1)} \int_0^T \E \sum_{i=1}^d \left|
                         \summ{j} \sum_{k, \ell = 1}^d
          \frac{f_{ij,k \ell}(x(s)) y_k(s) y_{\ell}(s) \eta_{j}(s)}
          {\alpha_j} \right|^{2p}  \, ds
             \nonumber \\ & \leq &
           C \epsilon^{2p(\gamma + 1)} \int_0^T \E \left( \|y(s) \|^{4p} \left( \summ{j}
           \alpha_j^{\alpha + \gamma - 1} \eta_j(s) \right)^{2p} \right) \, ds
            \nonumber \\ & \leq &
            C \epsilon^{2p(\gamma - 1) - \sigma},
\nonumber
\end{eqnarray}
where $\sigma > 0$.

Consider now $H_7(t)$. This term can be written in the following form:
\begin{equation}
H_7(t) = \epsilon \int_0^t df(x(s)) b(s) A^{-1} \eta(s) \, ds.
\nonumber
\end{equation}
We use the H\"{o}lder inequality, together with a calculation similar to the one
presented in the proof of Lemma \ref{lem:y_int} as well as assumptions
\eqref{e:b_cond}, \eqref{eqn:conds_g_s_1} with $\rho=1$ to obtain:
\begin{eqnarray}
\nrm{H_7(t)}{p} & \leq & C \epsilon^{2p} \int_0^T \E \left\| df(x(s)) b(x(s))
                                A^{-1}  \eta(s) \right\|^{2p} \, ds
            \nonumber \\  & \leq &
             C \, \epsilon^{2p} \|b(x)\|^{2p}_{L^{\infty}(\R^d)} \int_0^T \E \left( \summ{j}
         \alpha^{\beta + r - 1} |\eta_j(s)| \right)^{2p} \, ds
            \nonumber \\ & \leq &   C \, \epsilon^{2p}.
\end{eqnarray}
Consider finally $H_8(t)$. We write it in the form
$$
H_8(t) = \epsilon^{\gamma} \int_0^t df(x(s)) y(s) A^{-1} dW(s).
$$
We use the Burkholder--Davis--Gundy inequality to obtain
$$
\nrm{H_8(t)}{p} \leq C \epsilon^{2 \gamma p} \E \left( \int_0^T \| df(x(s)) y(s)
A^{-1} \|^2_{L^0_2} \right)^p.
$$
Now we have
$$
\| df(x(s)) y(s) A^{-1} \|^2_{L^0_2} \leq C \|y(s) \|^2,
$$
provided that condition \eqref{eqn:conds_g_s_2} with $\rho = 0$ holds. Thus:
$$
\nrm{H_8(t)}{p} \leq C \epsilon^{2 \gamma p} \E \|y(t) \|^{2p} \leq C
\epsilon^{2p(\gamma-1) -\sigma}.
$$
Putting now all the above estimates together we obtain \eqref{e:h_app}. \qquad\end{proof} }
\section*{Acknowledgments} The authors are grateful to D. Cai, P.R. Kramer and J.C.
Mattingly for useful suggestions. They are particularly grateful to J.M. Sancho for
useful suggestions and for providing them with references \cite{sancho2, sancho1}.

\def\cprime{$'$} \def\Rom#1{\uppercase\expandafter{\romannumeral
  #1}}\def\u#1{{\accent"15 #1}}\def\Rom#1{\uppercase\expandafter{\romannumeral
  #1}}\def\u#1{{\accent"15 #1}}


\begin{thebibliography}{99}

\bibitem{adler}
R.~J. Adler.
\newblock {\em An {I}ntroduction to {C}ontinuity, {E}xtrema, and {R}elated
  {T}opics for {G}eneral {G}aussian {P}rocesses}, volume~12 of {\em Institute
  of Mathematical Statistics Lecture Notes---Monograph Series}.
\newblock Institute of Mathematical Statistics, Hayward, CA, 1990.

\bibitem{arnold_l}
L.~Arnold.
\newblock {\em Stochastic differential equations: theory and applications}.
\newblock Wiley-Interscience [John Wiley \& Sons], New York, 1974.
\newblock Translated from the German.

\bibitem{aurich}
R.~Aurich, A.~B{\"a}cker, R.~Schubert, and M.~Taglieber.
\newblock Maximum norms of chaotic quantum eigenstates and random waves.
\newblock {\em Phys. D}, 129(1-2):1--14, 1999.

\bibitem{blakenship}
G.~Blankenship and G.C. Papanicolaou.
\newblock Stability and control of stochastic systems with wide-band noise
  disturbances. {I}.
\newblock {\em SIAM J. Appl. Math.}, 34(3):437--476, 1978.

\bibitem{bouc}
R.~Bouc and {\'E}.~Pardoux.
\newblock Asymptotic analysis of {PDE}s with wide-band noise disturbances, and
  expansion of the moments.
\newblock {\em Stochastic Anal. Appl.}, 2(4):369--422, 1984.

\bibitem{carmona_2}
R.~A. Carmona and J-P. Fouque.
\newblock Diffusion-approximation for the advection-diffusion of a passive
  scalar by a space-time {G}aussian velocity field.
\newblock In {\em Seminar on Stochastic Analysis, Random Fields and
  Applications (Ascona, 1993)}, volume~36 of {\em Progr. Probab.}, pages
  37--49. Birkh\"auser, Basel, 1995.

\bibitem{carmona}
R.A. Carmona and L.~Xu.
\newblock Homogenization theory for time-dependent two-dimensional
  incompressible gaussian flows.
\newblock {\em The Annals of Applied Probability}, 7(1):265--279, 1997.

\bibitem{dowell}
R.M. Dowell.
\newblock {\em Differentiable Approximation to Brownian Motion on Manifolds,
  Ph.D Thesis}.
\newblock University of Warwick, Coventry, UK, 1980.

\bibitem{ethier_86}
S.N. Ethier and T.G. Kurtz.
\newblock {\em Markov processes}.
\newblock Wiley Series in Probability and Mathematical Statistics: Probability
  and Mathematical Statistics. John Wiley \& Sons Inc., New York, 1986.

\bibitem{fannjiang}
A.~C. Fannjiang.
\newblock Convergence of passive scalar fields in {O}rnstein-{U}hlenbeck flows
  in {K}raichnan's model.
\newblock {\em J. Statist. Phys.}, 114(1-2):115--135, 2004.

\bibitem{sancho}
J.~Garc{\'{\i}}a-Ojalvo and J.~M. Sancho.
\newblock {\em Noise in {S}patially {E}xtended {S}ystems}.
\newblock Institute for Nonlinear Science. Springer-Verlag, New York, 1999.

\bibitem{raz}
D.~Givon and R.~Kupferman.
\newblock White noise limits for discrete dynamical systems driven by fast
  deterministic dynamics.
\newblock {\em Physica A}, 335:385--412, 2004.

\bibitem{graham}
R.~Graham and A.~Schenzle.
\newblock Stabilization by multiplicative noise.
\newblock {\em Phys. Rev. A}, 26(3):1676--1685, 1982.

\bibitem{grieser}
D.~Grieser.
\newblock Uniform bounds for eigenfunctions of the {L}aplacian on manifolds
  with boundary.
\newblock {\em Comm. Partial Differential Equations}, 27(7-8):1283--1299, 2002.

\bibitem{lefever}
W.~Horsthemke and R.~Lefever.
\newblock {\em Noise-induced transitions}, volume~15 of {\em Springer Series in
  Synergetics}.
\newblock Springer-Verlag, Berlin, 1984.
\newblock Theory and applications in physics, chemistry, and biology.

\bibitem{kraichnan}
R.~H. Kraichnan.
\newblock Diffusion by a random velocity field.
\newblock {\em Phys. Fluids}, 13(1):22--31, 1970.

\bibitem{kramer_JSP}
P.R. Kramer.
\newblock Two different rapid decorrelation in time limits for turbulent
  diffusion.
\newblock {\em J. Statist. Phys.}, 110(1-2):83--136, 2003.

\bibitem{paper3_stuart}
R.~Kupferman, G.~A. Pavliotis, and A.M. Stuart.
\newblock It\^{o} versus {S}tratonovich white noise limits for systems with
  inertia and colored multiplicative noise.
\newblock {\em Phys. Rev. E}, 70:036120, 2004.

\bibitem{kushner1}
H.~J. Kushner and H.~Huang.
\newblock Approximating multiple {I}t\^o integrals with ``band limited''
  processes.
\newblock {\em Stochastics}, 14(2):85--113, 1985.

\bibitem{kushner2}
H.~J. Kushner and H.~Huang.
\newblock Limits for parabolic partial differential equations with wide band
  stochastic coefficients and an application to filtering theory.
\newblock {\em Stochastics}, 14(2):115--148, 1985.

\bibitem{mangioni}
S.~E. Mangioni, R.~R. Deza, R.~Toral, and H.~S. Wio.
\newblock Nonequilibrium phase transitions induced by multiplicative noise:
  Effects of self--correlation.
\newblock {\em Phys. Rev. E}, 61(1):223--232, 2000.

\bibitem{nelson}
E.~Nelson.
\newblock {\em Dynamical theories of {B}rownian motion}.
\newblock Princeton University Press, Princeton, N.J., 1967.

\bibitem{oksendal}
B.~Oksendal.
\newblock {\em Stochastic Differential Equations}.
\newblock Springer-Verlag, Berlin-Heidelberg-New York, 1998.

\bibitem{paper1_stuart}
G.~A. Pavliotis and A.~M. Stuart.
\newblock White noise limits for inertial particles in a random field.
\newblock {\em Multiscale Model. Simul.}, 1(4):527--533 (electronic), 2003.

\bibitem{prato92}
G.~Da Prato and J.~Zabczyk.
\newblock {\em Stochastic {E}quations in {I}nfinite {D}imensions}, volume~44 of
  {\em Encyclopedia of Mathematics and its Applications}.
\newblock Cambridge University Press, 1992.

\bibitem{reimann}
P.~Reimann.
\newblock Brownian motors: noisy transport far from equilibrium.
\newblock {\em Phys. Rep.}, 361(2-4):57--265, 2002.

\bibitem{sancho2}
J.~M. Sancho, M.~San~Miguel, and D.~D{\"u}rr.
\newblock Adiabatic elimination for systems of {B}rownian particles with
  nonconstant damping coefficients.
\newblock {\em J. Statist. Phys.}, 28(2):291--305, 1982.

\bibitem{sancho1}
J.M. Sancho and A.~Sanchez.
\newblock External fluctuations in front dynamics with inertia: the overdamped
  limit.
\newblock {\em Eur. Phys. J. B}, 16:127--131, 2000.

\bibitem{inertial_1}
H.~Sigurgeirsson and A.~M. Stuart.
\newblock Inertial particles in a random field.
\newblock {\em Stoch. Dyn.}, 2(2):295--310, 2002.

\bibitem{inertial_2}
H.~Sigurgeirsson and A.~M. Stuart.
\newblock A model for preferential concentration.
\newblock {\em Phys. Fluids}, 14(12):4352--4361, 2002.

\bibitem{sogge}
C.~D. Sogge.
\newblock {\em Fourier integrals in classical analysis}, volume 105 of {\em
  Cambridge Tracts in Mathematics}.
\newblock Cambridge University Press, Cambridge, 1993.

\bibitem{wong}
E.~Wong and M.~Zakai.
\newblock On the convergence of ordinary integrals to stochastic integrals.
\newblock {\em Ann. Math. Statist.}, 36:1560--1564, 1965.

\bibitem{zabczyk}
J.~Zabczyk and G.~Tessitore.
\newblock Wong--{Z}akai approximations of stochastic evolution equations.
\newblock {\em Warwick Preprint}, 9, 2001.
\end{thebibliography}
\end{document}